\documentstyle[preprint,eqsecnum,aps]{revtex}

\newcommand{\be}{\begin{equation}}
\newcommand{\ee}{\end{equation}}
\newcommand{\bea}{\begin{eqnarray}}
\newcommand{\eea}{\end{eqnarray}}

\renewcommand{\baselinestretch}{1.0}

\begin{document}

\preprint{HD-TVP-9702}

\title{\vspace{1.5cm} Derivation of  transport equations for a 
strongly interacting Lagrangian in powers of 
$\hbar$ and $1/N_c$.\footnote{Dedicated to Professor R.H. Lemmer on the 
occasion of his 65th birthday}
      }

\author{S.P.~Klevansky, A.~Ogura and J.~H\"ufner }
\address{ Institut f\"ur Theoretische Physik, \\
Philosophenweg 19, D-69120 Heidelberg, Germany}

\maketitle
\vspace{1cm}

\begin{abstract}
Transport theory for an interacting fermionic system is reviewed and
applied to the chiral Lagrangian of the Nambu--Jona-Lasinio model.
Two expansions must be applied: an expansion in the inverse number
of colors, $1/N_c$, due to the nature
of the strong coupling theory, and a semiclassical expansion,
in powers of $\hbar$.
The quasiparticle approximation is implemented at an ear\-ly stage, and 
spin effects are omitted.
The self-energy is evaluated,  self-consistently only in 
the Hartree approximation, and semi-perturbatively in the collision
integral. 
   In the Hartree approximation, $O((1/N_c)^0)$,  the Vlasov
equation is recovered to $O(\hbar^1)$, 
together with an on-mass shell constraint equation, that is automatically
fulfilled by the quasiparticle ansatz.    The expressions for the
self-energy to order $O((1/N_c))$ lead to the collision term.
    Here one sees explicitly that
particle-antiparticle creation and annihilation processes are suppressed
that would otherwise be present, should an off-shell energy spectral
function be admitted.
A clear identification of the $s$, $t$ and $u$ channel scattering processes
in connection with the self-energy graphs is made and the origin of the
mixed terms is made evident.   Finally, after ordering according to powers 
in $\hbar$,
a  Boltzmann-like form
for the collision integral is obtained.
\end{abstract}

\clearpage
\renewcommand{\baselinestretch}{1.3}
\footnotesize\normalsize

\section{Introduction}

The description of physical systems that are out of thermal and chemical
equilibrium has its origins in  the early work of Boltzmann in 
1872 which describes the development of a system via an equation for
the single particle distribution function.   The formal derivation
   of non-relativistic 
transport equations from a specific field
theory can be traced  back to the work of Kadanoff and Baym \cite{kaba},
but which can be succinctly stated in a closed time path (CTP) formalism
that is independently due to Schwinger and Keldysh \cite{keldysh}
\footnote{Some work that already introduced the necessary concepts  
by Bakshi and Mahanthappa was already published in Ref.\cite{bama}}.
To date, this formalism has been applied to many non-relativistic problems,
but also to some in its relativistic generalization.   There are several
excellent books and reviews available \cite{degroot,landau,malfliet,chou},
which elucidate the techniques, both from an operator or path integral
point of view, to which we refer the reader.

Recently there has been an additional surge of interest in non-equilibrium
formulations of field theory from the nuclear physics community, as a need
exists for describing heavy-ion collisions within such a framework.   
A seminal article on this topic is that of Mr\'owczy\'nski and Heinz
\cite{heinz}, who, independently of  Sch\"onhofen et al. \cite{schoen}, have
derived transport
equations for the Walecka model. 
The derivation of such transport equations is in fact more general than 
appears on first sight, and the basic equations  of motion for the
Green functions, as they occur in Ref.\cite{heinz} are applicable to any field
theory.    The major difficulty in the fermionic
sector lies in the fact that
(a) there are in general 16 $\times$ 2 equations to be solved, and
(b) should one in any event introduce  a simplifying assumption such as
the quasiparticle approximation, one has to perform a complicated evaluation
of the self-energy that goes beyond the mean field approximation, if
one wishes to describe interesting physics.  
While this has often been done (see for example, Ref.\cite{schoen,perry}), it
is often unclear how the final collision integral is related to
scattering diagrams {\it in a strict Feynman sense}, although one knows
heuristically that the Boltzmann equation must emerge from a semi-classical
approximation.

The purpose of this paper is twofold.   On the one hand, we wish to make
clear how the collision term in a specific model is related to the 
Boltzmann equation via Feynman (and {\it not} heuristic) diagrams in
evaluating the scattering cross section that appears in this expression,
 and also to
explicitly demonstrate the limits of the quasiparticle assumption in that
it supresses vacuum fluctuations and particle production and annihilation
processes that would be present if the on-shell condition were relaxed.
Although the calculation is performed for a specific model, its generality
is
evident.   A transport calculation for QCD, or any other theory involving
relativistic fermions, for that matter, should also display the structure
given in this paper\footnote{This is not evident, for example, in the
diagrams shown by Geiger \cite{geiger} who omits exchange graphs.}
Thus this calculation gives a working guide for the general construction of
the collision term.

The second purpose of this manuscript is to develop the transport theory for
a fermionic {\it chiral} Lagrangian, in our case, the minimal version of the
Nambu--Jona-Lasinio (NJL) model \cite{njl,reviews}.
   This specific choice is motivated by the
following physical considerations.    Lattice simulations of QCD, which are
{\it equilibrium} studies of this field theory, show a chiral phase
transition at  a fixed value of the temperature \cite{laerman}, a feature
which can be simply modelled by the NJL Lagrangian, see for example 
\cite{weise}.     Unfortunately, due to the difficulties inherent
in QCD itself, there appears to be no means of observing such a phase
transition in an equilibrium system 
 and extract, for example, critical exponents.
Rather, the experimental search for clues to a phase transition that has been
conceived, is to proceed via heavy ion collisions, in which both high
temperatures and densities are to be reached.   Much effort has been
expended in this direction and will continue to be so during the early
course of the next century, particularly at RHIC and CERN.   
In addition, the simple equilibrium picture of the static
lattice simulations is inadequate for a complete description of such
 heavy ion experiments.   We note that, from the theoretical side, there is
at present no single conclusive
observable that might indicate the occurrence of the chiral phase
transition, given that one is dealing with a non-equilibrium system\footnote{
Some research has been undertaken on the basis
of static calculations that imply that substantial changes my occur in the
mesonic spectrum, in particular with regard to the $\rho$ mass.    Such
changes, however, are inferred from model calculations, usually together
with additional assumptions \cite{laerman},
 and they differ according to the approach
taken.}.
Thus we may speculate as to whether a concrete signal of this phase
transition can be obtained from a non-equilibrium formulation of the problem
or transport theory, 
for a simplified field theory 
{\it that is based on an underlying chiral Lagrangian}
and which therefor displays the same symmetry structure as QCD.
Here one may surmise whether phenomena such as critical
scattering\cite{mott},
 i.e.
divergence of the scattering cross-sections at the phase transition, may
leave an obvious signal, or simply the fact that the dynamically generated
quark mass is a function of space and time may leave some visible remnant
of a phase transition.

Our starting point is thus the Nambu--Jona-Lasinio (NJL) model, that
intrinsically
contains only quark degrees of freedom that interact via an effective two
body (in $SU_f(2)$) and, in addition, 
 three body interactions for three flavors. 
  Mesonic states are
constructed as collective excitations in this model.    No explicit gluonic
degrees of freedom are included.      
For our purposes, it is sufficient to study the simplest form of this model
that displays the features of chiral symmetry, and we will thus restrict
ourselves to the simplest form of Lagrange density that does so.

For the derivation of transport equations,
 the Schwinger-Keldysh formalism \cite{keldysh}
 for non-equilibrium
Green functions is applied, in what is by now a standard approach
\cite{landau,malfliet,chou}.    We follow closely the approach of
Ref.\cite{heinz}.
The associated equations of motion for the Wigner functions are then written
down in a general manner that pertains to {\it any} fermionic theory.  The 
difficulty in solution, on the other hand, as has been pointed out,
has its origins primarily in the
complication caused by the fact that one is dealing with relativistic 
spinors \cite{pfheinz}.

In this paper, we construct the transport and constraint equations for the
NJL model Lagrange density that we have chosen.   There are four esssential
ingredients to this calculation.

   (i)  We choose
 to make a quasiparticle expansion of the Wigner
function at a rather early stage of the derivation,
 and (ii) we do not allow for spin effects.   This reduces the
independent functions to simply be the distribution functions for quarks
and antiquarks for which equations of motion are derived, whose ingredients
are the self-energies.    These two assumptions, while not necessary in 
principle, reduce the complexity of the manipulations tremendously.   They
also correspond to what is at present possible in numerical calculations.   
(iii)  Noting that the NJL model is a strong interaction theory, one
requires a non-perturbative expansion of the self-energies
 in the inverse number of colors
$1/N_c$ \cite{quack,lemmer}. (iv)
In addition, transport equations are usually made tractable by introducing
a derivative expansion.    
We take the classical limit of the transport equation, classifying all the
terms in powers of $\hbar$.   Note that this only partly corresponds to 
the derivative expansion that is usually made.
   Thus there are {\it two} relevant expansions here, which are
to be
overlaid. 
We treat this double expansion with great care, and explicitly display all 
powers of $\hbar$ explicitly.
  It is essential at each stage of the expansion,  that in addition to the
usual constraint of energy conservation, chiral  symmetry is also preserved.
To this end, (i) the $1/N_c$ expansion must be consistently applied and
(ii)
it is important to see that the semiclassical expansion does not destroy the
symmetry properties.  This has been rigorously proven for the lowest order
terms in $1/N_c$.

  To start with, we examine the  terms of $O((1/N_c)^0)$
in the expansion.   This corresponds to taking the Hartree term
only. 
  Such a calculation is usually performed
self-consistently    and this is our approach  here also.
We then derive the
transport and constraint equations to $O((1/N_c)^0)$, and this is
done   
semi-classically, where only the leading  terms  according to powers of
$\hbar$ are retained.
Note that the aspect of chiral symmetry enters into these equations 
at this point in the property of the
 subsidiary equation, the gap equation, which  must be
solved concomitantly with the transport and constraint equations.

We next move to  terms of order $O((1/N_c)^1)$
 that give rise to collisions.
The lowest of these terms contains two interaction lines.
According to our philosophy of working in a $1/N_c$ expansion,
 the self-consistently evaluated Hartree Green functions must now be employed
in this evaluation.
In this approximation, we again derive the transport and constraint 
equations, and again use $\hbar$ ordering to decide to what level terms
in these equations should be accepted.   In doing so,
    we are able to demonstrate
explicitly, in a fully relativistic fashion, that the Boltzmann-like 
collision term is recovered, containing as heuristically expected, matrix
elements for the elastic scattering processes, $qq'\rightarrow qq'$ and
$q\bar q'\rightarrow q\bar q'$, here of course only in the Born
approximation.  
In an exact sense, we have rewritten the collision term in terms of
scattering cross-sections with {\it well defined Feynman graphs}, and we
have illustrated explicitly which self-energy graphs are required to 
obtain our result.
In doing this, we have generalized the 
non-relativistic treatment of Kadanoff and Baym \cite{kaba}.
Several other authors have examined the collision term, see for example
Refs.\cite{schoen,perry}.    To our knowledge, however, this is the first
paper that links the cross-sections that occur in the Boltzmann-like
equation with Feynman graphs for the scattering  
processes.

    There is, however, a 
major difference that arises here.   By direct calculation, one sees 
explicitly that, even at this level, terms which describe
 the three body creation and
annihilation processes, such as $q\leftrightarrow (q\bar q)q$
as well as vacuum fluctuations, are present, but 
are suppressed by the on-shell requirement of the
quasiparticle approximation.    Relaxing this condition would lead
automatically to particle production / destruction.

The above calculations have been carried out, in this first exploratory 
article, by considering only one term in the $1/N_c$ expansion.    Clearly the
full set is required that would completely include 
 the mesonic degrees of freedom.     This generalization to include the
complete mesonic sector is however in itself a complex problem, which builds on the
concepts and ideas presented here.   Again, in this context, care must be
taken in the $\hbar$ expansion, to see that the properties of chiral symmetry
remain intact.
  Using these 
methods, it should become 
possible to construct a consistent theory of quarks and
mesons in which the collision term contains {\it inter alia} the
hadronization processes $q\bar q'\rightarrow MM'$ in addition to the usual
elastic scattering $q\bar q'\rightarrow q\bar q'$ and $qq'\rightarrow qq'$
in the quark sector.    The details of the mesonic sector, however, are 
 left to a future publication 
\cite{ski}.

We conclude this section by mentioning some other papers that are relevant, but
we do not attempt to provide an exhaustive list.    Relativistic kinetic
equations for a Yukawa model were first derived in Ref.\cite{perry}.   Here
the difficulties in expressing the collision integral in terms of actual
Feynman graphs is apparent.   A seminal paper discussing the Walecka model
has followed this \cite{heinz}, see also \cite{schoen}.
   A transport theory for the NJL model,
derived from a path integral approach with a semi-bosonized Lagrangian
 was first undertaken by \cite{wilets}.
We differ from this in two respects, viz (i) our approach is canonical, 
and treats the theory from a fermionic point of view, and 
(ii) that the terms are ordered  according to the double expansion with 
regard to $1/N_c$ as well as the power counting in $\hbar$.

This paper is organized as follows: in Section II, we review the general
non-equilibrium transport theory for fermions, derived from a canonical
point
of view.   We construct the general transport and constraint equations.   In
Section III, we discuss the quasiparticle approximation, introduce
our Lagrange density and explain the strategy of the $1/N_c$ and $\hbar$
expansions.
In Section IV, we calculate the mean field self-energy and 
transport and constraint equations in the Hartree approximation.    We devote
Section V to the evaluation of the collision term
 and demonstrate that 
the classical limit leads us to the Boltzmann equation.    Off-shell terms
that are dropped are seen explicitly. 
  The constraint equation, to this level of
approximation, is also given.   
In this section, we present the major new work of this paper.
 We conclude in Section VI.   We have,
 for completeness, included several appendices that give closer
detail of some of the technical and / or repetitive aspects of the 
derivations.

\vskip 0.5in

\section{General transport theory for fermions}

\subsection{Equations of motion}
 
Non-equilibrium phenomena are completely described via the
Schwinger - Keldysh formalism for Green functions \cite{keldysh}.
   Many conventions
exist in the literature.   
In this section, we give ours.
For our purposes, it is simplest to use
the convention of Landau\footnote{
Note that this convention differs from that of Ref.\cite{malfliet} in that
those authors attribute the {\it opposite} signs to the contour
branches.   There is also another difference that arises in that the
off diagonal self-energies in this convention of \cite{landau} differ from
that of \cite{malfliet} and \cite{heinz} by a minus sign.} \cite{landau}.
   In this, the designations $+$ and
$-$ are
attributed to the closed time path that is shown in Fig.1, and the
fermionic Green functions are defined as
\begin{eqnarray}
i\hbar S^c(x,y) &=& \left <T\psi(x) \bar \psi(y)\right > = i\hbar S^{--}(x,y)\nonumber
\\
i\hbar S^a(x,y) &=& \left <\tilde T\psi(x)\bar\psi(y)\right > = i\hbar S^{++}(x,y)
\nonumber \\
i\hbar S^{>}(x,y) &=& \left <\psi(x)\bar\psi(y)\right > = i\hbar S^{+-}(x,y) \nonumber
\\
i\hbar S^<(x,y) &=& -\left <\bar\psi(y)\psi(x)\right > = i\hbar S^{-+}(x,y),
\label{gfunctions}
\end{eqnarray}
where $T$ is the usual time ordering operator and $\tilde T$ the {\it anti}-time
ordering operator.
These can be summarized into the compact matrix notation
  \begin{equation}\label{green}
    \underline{S} = \left( \begin{array}{cc}
                              S^{--} & S^{-+} \\
                              S^{+-} & S^{++}
                           \end{array} \right).  
  \end{equation}                   
The matrix of irreducible proper self-energies,
 \begin{equation}\label{self}                 
    \underline{\Sigma} = \left( \begin{array}{cc}
                                   \Sigma^{--} & \Sigma^{-+} \\
                                   \Sigma^{+-} & \Sigma^{++}
                                \end{array} \right),
  \end{equation}
is introduced via the Dyson equation,
 \begin{eqnarray}
    \underline{S}(x,y) &=& \underline{S^{0}}(x,y)  
                  + \int d^{4}z d^{4}w \underline{S^{0}}(x,w)  
                                        \underline{\Sigma}(w,z)
                                        \underline{S}(z,y)\nonumber \\
                    &=& \underline {S^{0}}(x,y)                          
                  + \int d^{4}z d^{4}w \underline{S}(x,w)            
                                        \underline{\Sigma}(w,z)          
                                        \underline{S^0}(z,y)
\label{dyson}
  \end{eqnarray}
where $\underline S^{0}$ is the matrix of propagators for the
non-interacting system.
In a standard fashion, one
constructs equations of motion for the matrix of Green functions.
These take the form
  \begin{equation}
    \left( i \hbar \not\!\partial_{x} -m_0 \right) \underline{S}(x,y) =
        \underline{ \sigma_{z} } \delta^{4} (x-y)
                  + \int d^{4}z  \underline{ \sigma_{z} }
                                        \underline{\Sigma}(x,z)
                                        \underline{S}(z,y) ,
 \label{eq:401}
 \end{equation}
where
  \begin{equation}
    \underline{ \sigma_{z} } = \left( \begin{array}{cc}
                           1 &  0 \\
                           0 & -1
                        \end{array}\right)
  \end{equation}
has been introduced.
Note that the entries in the matrices of Green functions, Eq.(\ref{green})
and self-energies Eq.(\ref{self}) are amongst one another not independent.
They are linearly related in a way that is obvious from their definitions,
i.e.
\begin{equation}                                         
S^{--}_{\alpha\beta} (x,y) + S^{++}_{\alpha\beta}(x,y) =
S^{-+}_{\alpha\beta} (x,y) + S^{+-}_{\alpha\beta}(x,y).
\label{linear}
\end{equation}
It is also useful to introduce the retarded and advanced Green functions,
$S^R$ and $S^A$ respectively, that are defined as
\begin{equation}
i\hbar  S^{\rm R}_{\alpha \beta} ( x,y ) 
              =  \theta(x_{0}-y_{0})
                 \langle \left\{ \psi_{\alpha}(x),
                                \overline{\psi}_{\beta}(y) \right\} \rangle
\label{retard}
\end{equation}
and
\begin{equation}
    i\hbar S^{\rm A}_{\alpha \beta} ( x,y )
               = - \theta(y_{0}-x_{0})
                 \langle \left\{ \psi_{\alpha}(x),
                                \overline{\psi}_{\beta}(y) \right\} \rangle.
\label{advance}
\end{equation}
Important relationships that connect the Green functions with each other, as
well as those pertaining to the self-energies, are listed in Appendix A. 
In
particular, it is also useful to note that 
\begin{equation}
\Sigma^R = \Sigma^{--} + \Sigma^{-+}
\label{sigret}
\end{equation}
and 
\begin{equation}
\Sigma^A = \Sigma^{--} + \Sigma^{+-}
\label{sigadv}
\end{equation}
link the retarded and advanced self-energies to $\Sigma^{ij}$, $i,j=\pm$.

\subsection{Transport and constraint equations.}

In this section, we give a brief derivation of the 
transport and constraint
equations following the lines of Ref.\cite{heinz}
\footnote{We will recover those equations 
of Ref.\cite{heinz} in the fermionic sector.
Here, however, we retain all factors of $\hbar$ explicitly.},
 and writing them in a 
concise form.
 It will be sufficient for us to examine only one component of this 
equation, the off diagonal element $S^{-+}$, together with the equation for
the hermitian conjugate.    
This simplification arises from the quasiparticle approximation (see below)
in which all four Green functions $S^{ij}$ can be expressed in terms of
two distribution functions.
The equations of motion  are
\begin{equation}
 \left( i \hbar \not\!\partial_{x} - m_0 \right)_{\alpha \beta}
              S^{-+}_{\beta \gamma}(x,y)     =\int d^{4} z \left\{
\Sigma^{--}_{\alpha \beta} ( x,z )
                               S^{-+}_{\beta \gamma}(z,y)
                            + \Sigma^{-+}_{\alpha \beta} ( x,z )
                               S^{++}_{\beta \gamma}(z,y) \right\}
\label{s}
\end{equation}
and
\begin{equation}
S^{-+}_{\alpha \beta}(x,y)
           \left( -i \hbar \overleftarrow{\not\!\partial_{y}}
                                 -m_0 \right)_{\beta \gamma}  =
 \int d^{4}z
        \left\{ - S^{--}_{\alpha \beta}(x,z)
                  \Sigma^{-+}_{\beta \gamma}(z,y)
                - S^{-+}_{\alpha \beta}(x,z)
                  \Sigma^{++}_{\beta \gamma}(z,y) \right\}
\label{scross}
\end{equation} 
It turns out to be expedient to eliminate $S^{++}$, $S^{--}$, $\Sigma^
{++}$ and $\Sigma^{--}$ in terms of the advanced / retarded forms,
together with the off-diagonal elements of the Green functions and
self-energies, as this will lead us directly to a convenient form for the
transport equation.   Employing Eqs.(\ref{eq:sr}), (\ref{eq:sa}) and
(\ref{tsr}) - (\ref{relationsig}), one can rewrite Eqs.(\ref{s})
and (\ref{scross}) as\footnote{Note that the more usual form of
the first equation would be 
\begin{equation}
\left(i\hbar\not\!\partial_x - m_0 \right)_{\alpha\beta}
S^{-+}_{\beta\gamma} (x,y) = \int d^4z \left \{\Sigma^R S^{-+} - \Sigma
^{-+} S^A \right \}, 
\end{equation}
from which one can immediately see that the right hand side is simply
the hermitian conjugate of the right hand side of Eq.(\ref{sagain}).    The
form that we have chosen, however, already anticipates the loss and gain
terms that one expects of the collision integral.}
  \begin{eqnarray}
     \left( i \hbar \not\!\partial_{x} - m_0 \right)_{\alpha \beta}
              S^{-+}_{\beta \gamma}(x,y)         
&=&\int d^{4} z \{ \Sigma^{-+}_{\alpha \beta} (x,z)  
         S^{+-}_{\beta \gamma} (z,y) 
      - \Sigma^{+-}_{\alpha \beta} (x,z)  
         S^{-+}_{\beta \gamma} (z,y)                    \nonumber \\
    & &  + \Sigma^{\rm A}_{\alpha \beta} (x,z)
            S^{-+}_{\beta \gamma} (z,y)
         - \Sigma^{-+}_{\alpha \beta} (x,z)
            S^{\rm R}_{\beta \gamma} (z,y)  \}
\label{s1again}
\end{eqnarray}
and
\begin{equation}
     S^{-+}_{\alpha \beta}(x,y)
           \left( -i \hbar \overleftarrow{\not\!\partial_{y}}
                                 -m_0 \right)_{\beta \gamma} 
 = \int d^{4}z
        \left\{ - S^{\rm R}_{\alpha \beta}(x,z)
                  \Sigma^{-+}_{\beta \gamma}(z,y)
                + S^{-+}_{\alpha \beta}(x,z)
                  \Sigma^{\rm A}_{\beta \gamma}(z,y) \right\},
\label{sagain}
  \end{equation}
One now 
moves to relative and centre of mass variables, $u=x-y$ and $X=(x+y)/2$.  A
Fourier transform with respect to the relative coordinate, or Wigner
transform, \begin{equation} S(X,p) = \int d^4u e^{ip\cdot u/\hbar}
 S\left(X+\frac
u2,X-\frac u2\right) 
\label{wig}
\end{equation} is then performed on the above 
two equations.  
 Simple rules for the calculation with the Wigner transform
 are given in Appendix B.
In general, the Wigner transforms of Eqs.(\ref{s1again}) and (\ref{sagain})
are 
\begin{eqnarray}
\{ i\hbar\gamma^\mu(\frac 12\frac{\partial}{\partial X^\mu}&-& \frac
{ip_\mu}\hbar) -m_0\} S^{-+}(X,p) = \nonumber \\
& &\Sigma^{-+}(X,p) \hat \Lambda
S^{+-}(X,p) - \Sigma^{+-}(X,p)\hat\Lambda S^{-+}(X,p) \nonumber \\ 
&+& \Sigma^{A}(X,p)\hat \Lambda S^{-+}(X,p) - \Sigma^{-+}
(X,p)\hat\Lambda S^R(X,p)
\end{eqnarray}
and
\begin{eqnarray}
S^{-+}(X,p)\{-i\hbar\gamma^\mu(\frac12\frac{\overleftarrow\partial}{
\partial X^\mu}
+ \frac{ip_\mu}\hbar) - m_0\} =
&-&S^{R}(X,p)\hat\Lambda \Sigma^{-+}(X,p) \nonumber \\ & +&
S^{-+}(X,p)\hat\Lambda\Sigma^{A}(X,p),
\end{eqnarray}
respectively, with
\begin{equation}
\hat\Lambda = \exp\left(-\frac{i\hbar}2(
\frac{\overleftarrow\partial}{\partial X^\mu}\frac
{\overrightarrow\partial}{\partial p_\mu} - \frac{
\overleftarrow\partial}{\partial p_\mu} \frac{\overrightarrow\partial}
{\partial X^\mu})\right).
\label{hatlambda}
\end{equation}
Now, on subtracting and adding the two resulting transformed
equations, one constructs 
the so-called transport and constraint equations:
\begin{equation}
 \frac{i\hbar}2\{\gamma^\mu,\frac{\partial
S^{-+}}{\partial X^\mu}\} + [\not p,S^{-+}(X,p)] =
I_{-}
\label{full1}
\end{equation} 
and
\begin{equation}
 \frac{i\hbar}2[\gamma^\mu,\frac{ \partial S^{-+}}{\partial
X^\mu}] + \{ \not p - m_0,S^{-+}\} = I_{+},
\label{full2}
\end{equation} 
respectively.  Here
\begin{equation}
I_{\mp} = I_{{\rm coll}} + I^A_{\mp} + I^R_{\mp},
\label{is}
\end{equation}
where the collision term is
\begin{eqnarray}
I_{{\rm coll}} &=&
\Sigma^{-+}(X,p) \hat\Lambda S^{+-}(X,p) - \Sigma^{+-}(X,p) \hat\Lambda
S^{-+}(X,p) \nonumber \\
&=& I_{{\rm coll}}^{{\rm gain}} - I_{{\rm coll}}^{{\rm loss}},
\label{collision}
\end{eqnarray}
and terms containing retarded and advanced components are contained
in
\begin{equation}
I^R_{\mp} =  - \Sigma^{-+} ( X,p ) \hat \Lambda
                                  S^{\rm R} ( X,p )
         \pm S^{\rm R} ( X,p ) \hat\Lambda
                                         \Sigma^{-+} ( X,p )
\label{irmp}
\end{equation}
and
\begin{equation}
I^A_{\mp} =\Sigma^{\rm A} ( X,p ) \hat\Lambda
                                  S^{-+} ( X,p )
         \mp S^{-+} ( X,p ) \hat\Lambda
                                      \Sigma^{\rm A} ( X,p ).
\label{iamp}
\end{equation}
Note that the equations (\ref{full1}) and (\ref{full2}) are
general equations that are generic for {\it any}
relativistic fermionic theory.     These are already well known, and
have been derived for the Walecka model in Ref.\cite{heinz}
and rederived for QCD in Ref.\cite{geiger}, for example. 

  At this point,
one has
  to specify the particular theory that one
wishes to investigate -- in our case, this is
 the simplest version of the NJL model
--
and to calculate the self-energy terms systematically.   
In particular, the difficulty that
we wish to address in this paper, is the evaluation of the collision term in
a form that is understandable on physical grounds.

\section{Approximation schemes and Feynman-like rules}

\subsection{Quasiparticle Approximation}

Equations (\ref{full1}) and (\ref{full2}) are each a system of coupled
equations for the 16 matrix elements of $S_{\alpha\beta}^{-+}$ in spinor
space.
A complete method of solution of these equations can be 
developed on inserting the spinor decomposition 
\begin{equation}
-i\hbar S^{-+} = F + i\gamma_5 P + \gamma^\mu V_\mu + \gamma^\mu\gamma_5 A_\mu + \frac
12\sigma^{\mu\nu}S_{\mu\nu},
\label{dec}
\end{equation}
for $S^{-+}$ into the above equation.
Similar decompositions for all other quantities are required, so that an
extremely complicated system of coupled equations results.
While this guarantees a correct solution from a {\it formal} point of view,
even in the collisionless system, it leads to a set of $16 \times 2$
equations that must be solved simultaneously \cite{vasak,woitek}.
   Only in limited cases
where spatial gradients have been neglected, can such a set of equations
been solved exactly numerically \cite{pfheinz}. 

A tremendous simplification arises from  the quasi-particle
approximation.       We do this in a fashion which, as will become evident
in the following section, represents a self-consistent solution within a
mean field approximation, i.e. we assume that the system consists of 
fermions that, in interacting only with the mean field, acquire a mass
$m(X)$ that depends on space and time,
and which will be dynamically generated due to the breaking of the chiral
symmetry of the underlying Lagrangian.
This is as yet undetermined.
The dressed fermions (the ``constituent quarks'') may then collide with each
other.   
The index ``$H$'' appearing on the Green functions below indicates that they
are the mean field (``Hartree'') functions.  
\begin{eqnarray}
S_H^{-+}(X,p) &=& 2\pi i \frac 1{2E_p}[\delta(p_0 -E_p)\sum_{ss'}u_{s'}(p)
\bar u_s(p) f^{ss'}_q(X,p) \nonumber \\
 &+& \delta(p_0+E_p) \sum_{ss'} v_{s'}(-p)\bar
v_s(-p)\bar f_{\bar q}^{ss'}(X,-p)],
\label{decomp}
\end{eqnarray}
and
\begin{eqnarray}
S_H^{+-}(X,p) &=& -2\pi i\frac 1{2E_p}[\delta(p_0-E_p)\sum_{ss '}u_s(p)\bar
u_s(p) \bar f_q^{ss'}(X,p)  \nonumber \\
&+& \delta(p_0+E_p)\sum_{ss'} v_s(-p)\bar v_{s'}(-p) f_{\bar
q}^{ss'}(X,-p)],
\label{decomp2}
\end{eqnarray}
that introduces the quark and antiquark distribution functions
$f_q	^{ss'}(X,p)$ and $f_{\bar q}^{ss'}(X,p)$ respectively.
These functions give the number density per degree of freedom in phase
space for on the energy shell quarks, $E^2_p = m^2(X) + \vec p^2$.
   In each case,
we have introduced the abbreviation $\bar f_{q,\bar q}^{ss'}(X,p) = 1 -
f_{q,\bar q}^{ss'}(X,p)$.    The $u$'s and $v$'s are normalized according
to $\bar u_s(p) u_s(p) = 2m\delta_{ss'}$ and $\bar v_s(p)v_s(p) = -2m\delta
_{ss'}$.   In principle, all spin combinations are possible.  We shall 
restrict ourselves, however, to the case $f_{q,\bar q}^{ss'} = \delta_{ss'}
f_{q,\bar q}$, i.e. we do not allow for spin polarizations. 
Under this assumption, one has
\begin{equation}
S_H^{-+}(X,p) = 2\pi i\frac {\not p + m}{2E_p}[\delta(p_0 - E_p)f_q(X,p) -
 \delta(p_0 + E_p)\bar f_{\bar q}(X,-p)]
\label{smp}
\end{equation}
and
\begin{equation}
S_H^{+-}(X,p) = - 2\pi i \frac{\not p + m}{2E_p}[\delta(p_0 - E_p) \bar f_q(X,p)
-\delta(p_0+E_p)f_{\bar q}(X,-p)] 
\label{spm}
\end{equation}
using $\sum_s u_s(p)\bar u_s(p) = \not p + m$ and $\sum_s v_s(p) \bar v_s
(p) = \not p - m$.
  Using this convention, the time ordered and anti-time ordered
free particle Green functions can be written as
\begin{eqnarray}
S_H^{--}(X,p) &=& \frac{\not p + m}{p^2-m^2+i\epsilon} \nonumber \\
& &  + 2\pi i \frac{\not p +
m}{2E_p}[\delta(p_0 - E_p) f_q(X,p) + \delta(p_0 + E_p) f_{\bar q}(X,-p)]
\label{smm}
\end{eqnarray}
and
\begin{eqnarray}
S_H^{++}(X,p) &=& - \frac{\not p + m}{p^2 - m^2-i\epsilon} \nonumber \\
& & + 2\pi i\frac
{\not p + m}{2E_p}[\delta(p_0-E_p) f_q(X,p) + \delta(p_0 + E_p) f_{\bar
q}(X,-p)].
\label{spp}
\end{eqnarray}
 Note that in  terms   
of the decomposition Eq.(\ref{dec}), the decomposition for $
S_H^{-+}$ above implies that only the scalar
and vector components,
\begin{equation}
F=-i\frac 14\hbar {\rm tr} S_H^{-+} = 2\pi \hbar \frac m{2E_p}[\delta(p_0 - E_p) f_q(X,p)
 - \delta(p_0+E_p) \bar f_{\bar q}(X,-p)]
\label{scalar}
\end{equation}
and
\begin{equation}
V^\mu = \frac{p^\mu}m F
\label{vector}
\end{equation}
survive.    Here the trace is regarded as being over the spinor variable 
only.  

  Since all four Green functions $S^{ij}(X,p)$ depend only on the two
distribution functions $f_q(X,p)$ and $f_{\bar q}(X,p)$, one needs only
to consider the equation for $S^{-+}$, c.f. Eq.~(\ref{s}).
Thus the $2\times 16$ equations reduce to two.

\subsection{The Lagrangian}

In order to proceed further, we must specify our Lagrange density.   
We choose the NJL model form as being
\begin{equation}
  {\cal L} =  \overline{\psi} \left( i \hbar \not\!\partial -
                                                    m_0 \right) \psi
            + G \left[ \left( \overline{\psi} \psi \right)^{2}
                 + \left( \overline{\psi} i \gamma^{5} \psi \right)^{2}
                  \right]~,
\label{lagrange}
\end{equation}
where $G$ is a dimensionful coupling, $[G]=$MeV$^{-2}$.  
Since the Lagrangian generates a non-renormalizable theory,
Eq.(\ref{lagrange}) has to be supplemented by a regularization scheme.
We shall use a $3-d$ cutoff $\Lambda$.
 For simplicity,
we have ignored the  flavor degree of freedom in Eq.(\ref{lagrange}).
   The color degree of 
freedom is to be implicitly understood.   This minimal form of the NJL model
is chirally symmetric for $m_0=0$, and exhibits dynamical symmetry breaking
at temperature $T=0$.     An equilibrium study of the model indicates that
a phase transition to a chirally symmetric state occurs at a finite value of
the temperature.    By formulating the non-equilibrium theory for this
model, we hope to gain insight into the {\it dynamical} consequences of chiral
symmetry breaking and restoration.  
The basic problems concerning formulation of the transport theory can be
studied within the framework of this simpler model, and the generalizations
to include isospin are obvious. 

    From a perturbative expansion in the interaction picture, 
Feynman-like rules can be constructed for the component functions of the
matrix of Green functions and their self-energies.   These follow on 
considering the same skeleton graphs that one would derive for the time
ordered equilibrium Green function using an extended interaction, and 
decorating the interaction line with the same sign at each vertex.    To 
each interaction line with a $+$ or $-$ sign is attributed a function 
$\mp \Gamma 2iG/\hbar \Gamma$, with $\Gamma= 1$ or $i\gamma_5$ for the scalar
or pseudoscalar vertex, as is required.    External legs have a sign 
attributed to them as well, in accordance with the function to be
evaluated.   Internal lines take all possible combinations of
the signs.  For example, the Dyson equation for the
full component $S^{++}(X,p)$
is given diagrammatically in Fig.~2.  
One can see that this diagram and the convention given are in accordance with 
Eq.(\ref{dyson}).
Further examples follow in the 
evaluation of the self-energy in the following sections.   

\subsection{The double expansion in powers of $1/N_c$ and $\hbar$}

Before we evaluate the right hand sides of the transport and constraint
equations, $I_-$ and $I_+$ respectively, we explain our strategy of a 
double expansion in powers of $1/N_c$ and $\hbar$.

In general, an expansion in powers of the coupling constant is inappropriate
for a strongly interacting theory.   In our theory, Eq.(\ref{lagrange}), the
relevant parameter is $G\Lambda^2$, which has a value close to two, and thus
precludes a perturbative expansion.   It is therefore customary to perform
suitable partial summations of relevant diagrams.     A perturbative
expansion in the inverse number of colors,
 $1/N_c$ has been proposed by Witten
in connection with QCD proper, and such an expansion technique
 has proven extremely useful
in connection with calculating observables in the NJL model.   In this
scheme, a diagram containing $n$ interaction vertices and $m$ fermion loops
is of order $G^nN_c^m = (N_c)^{n-m}$, if one sets $GN_c=1$, using a
symbolic notation.    Applied to the
NJL model, the partial summation at the level $O((1/N_c)^0)$ is equivalent
to the Hartree approximation (Fig.~3), and leads to the concept of
constituent quarks, while to order $O((1/N_c)^1)$, one recovers the 
Random Phase approximation which gives rise to mesonic masses and
$qq$ and $q\bar q$ cross sections.

When investigating the transition of a quantum system to its classical
limit, one often treats $\hbar$ as a small quantity (although this is 
inappropriate because of its dimension) and keeps only the leading orders in
$\hbar$ or takes the limit $\hbar\rightarrow 0$.    In the  transport and
constraint equations, Eqs.(\ref{full1}) and (\ref{full2}), the right
hand sides $I_-$ and $I_+$, respectively depend  on $\hbar$ in two ways:
(i) The derivative operator $\hat \Lambda$, Eq.(\ref{hatlambda}), contains
$\hbar$ to all orders.    Its expansion in $\hbar$ generates the various
terms of the so-called gradient expansion, and (ii) the self-energies
$\Sigma^R$ and $\Sigma^{-+}$ may have overall factors in $\hbar$.    While
to order $O((1/N_c)^0)$, $\Sigma^R = m$, and is directly related to the
mass, which we treat as a quantity that has a direct classical
interpretation, we find $\Sigma^{-+}\propto\hbar d\sigma/d\Omega$, and is only 
related via a factor $\hbar$ to a quantity with direct physical
interpretation.   Note that while we use the overall factors in $\hbar^0$ or
$\hbar^1$ in $\Sigma^R$ or $\Sigma^{-+}$ in our $\hbar$ classification, 
we do not expand the quantities $m$ or $d\sigma/ d\Omega$ in powers of 
$\hbar$.    We will detail this strategy in the calculation of the mean
field term and collision integrals.

\section{Mean Field Approximation}\label{harenergy}

   The first term in  the expansion of the self-energy in powers of $1/N_c$ is 
 the Hartree
diagram of Fig.~3.   This is evaluated self-consistently.
  Only diagonal elements exist for $\underline\Sigma_H$,
i.e. $\Sigma_H^{-+}(x,y) = \Sigma_H^{+-}(x,y) = 0$.     One finds by direct
calculation that
\begin{equation}
-i\Sigma_H^{--}(x,y) = -2i\hbar G\left({\rm tr} iS_H^{--}(x,x) + i\gamma_5 {\rm tr}
iS_H^{--}(x,x) i\gamma_5\right)\delta^4(x-y)
\label{hartreemm}
\end{equation}
while
\begin{equation}
-i\Sigma_H^{++}(x,y) = 2i\hbar G\left
({\rm tr} iS_H^{++}(x,x) + i\gamma_5 {\rm tr} 
iS_H^{++}(x,x) i\gamma_5\right)\delta^4(x-y),
\label{hartreepp}
\end{equation}
where ${\rm tr} = {\rm tr}_c{\rm tr}_\gamma = N_c{\rm tr}_\gamma$.
From this self-energy, we may derive two things, (i) the gap equation and
(ii) the transport and constraint equations at this level.  

Firstly, by examining the form of the retarded Green function for a free
particle, see Appendix A, one may identify the self-consistently determined
 mass of the particle to be
$m(X) = m_0+ \Sigma_H^R(X,\vec p=0) = m_0 + \Sigma_H^{--}(X,\vec p=0)$ to this level of
approximation.  Because of the $\delta$ functions in Eq.(\ref{hartreemm}),
$\Sigma_H^R(X)$ is in fact
 independent
of the momentum variable.
  Note that  $\Sigma_H^{--}$ and
$\Sigma_H
^{++}$ are scalars  in spinor space, since the second term in both
Eqs.(\ref{hartreemm}) and (\ref{hartreepp}) vanishes, in view of the
quasiparticle approximation Eq.(\ref{smp}).
Now, inserting the quasiparticle decomposition for $S_H^{--}(X,p)$,
see Eq.(\ref{smm}), with $m=m(X)$ into Eq.~(\ref{hartreemm})
 and performing a Wigner transform,
 we arrive at the gap equation for a non-equilibrium medium,
\begin{equation}
m(X) = m_0 + 4GN_cm(X)\int\frac{d^3p}{(2\pi\hbar)^3} \frac 1{E_p(X)} [ 1 - f_q(X,p)
- f_{\bar q}(X,p)],
\label{gap}
\end{equation}
 which is a non-linear equation for $m(X)$, and one 
recovers the result of \cite{wilets,woitek}.

We may now return to the transport and constraint equations and investigate
their form.   The only contribution from the right hand side of the 
transport and constraint equations, Eqs.(\ref{full1}) and (\ref{full2})
 that is non-vanishing, stems from the
advanced self-energy $\Sigma_H^A$ that occurs in $I^A_{\mp}$ in
Eq.(\ref{iamp}).     Closer
examination of $I^A_{\mp}$, shows that
\begin{equation}
I^A_{-} = -\frac{i\hbar}2\{\partial_\mu\Sigma_H^A(X), \partial^\mu_p S_H^{-+}(X,p)
\} 
\end{equation}
while
\begin{equation}
I^A_+ = \{\Sigma_H^A(X),S_H^{-+}(X,p) \}.
\end{equation} 
Here we have made use of the fact that $\Sigma^A$ is a scalar and is only
a function of $X$, and have
expanded $\hat\Lambda$ in Eq.(\ref{hatlambda}) to retain only the first 
non-vanishing term.
One thus has
\begin{equation}
\frac {i\hbar}2\{\gamma^\mu,\frac{\partial S_H^{-+}(X,p)}{\partial X^\mu}\} 
+[\not p,S_H^{-+}(X,p)] =  - i\hbar \partial_\mu \Sigma_H^A(X)
\partial_p^\mu S_H^{-+}(X,p)
\label{hartreetransport}
\end{equation}
and
\begin{equation}
\frac{i\hbar}2[\gamma^\mu,\frac{\partial S_H^{-+}}{\partial X^\mu} ]
+ \{\not p - m_0,S_H^{-+} \}= \{\Sigma_H^A(X), S_H^{-+}(X,p)\}
\label{hartreeconstraint}
\end{equation}
for the transport and constraint equations.   From these, one constructs 
equations for the distribution functions $f_{q,\bar q}(X,p)$ by inserting
Eq.(\ref{smp}) into (\ref{hartreetransport}) and (\ref{hartreeconstraint}), 
integrating over $p_0$ over  an interval $\Delta_{\pm}$ that contains $E_p(X)$ or
$-E_p(X)$ respectively, and performing a spinor trace.    Then Eq.
(\ref{hartreetransport}) becomes the Vlasov equation for $f_{q,\bar q}$,
\begin{equation}
p^\mu\partial_\mu f_{q,\bar q}(X,\vec p) + m(X)\partial_\mu m(X)\partial_p^\mu
f_{q,\bar q}(X,\vec p) = 0,
\label{vlasov}
\end{equation}
where $p^0=E_p(X)$.    We have detailed this derivation in Appendix C, since
some care must be taken in bringing the equation into this form.
The constraint equation, Eq.(\ref{hartreeconstraint}), 
which is somewhat simpler to treat, is also discussed briefly in Appendix C.
On inserting $S^{-+}
(X,p)$, and repeating the procedure described above, one finds
\begin{equation}
(p^2 - m^2(X))f_{q,\bar q} = 0.
\label{hatconstr}
\end{equation}
It is interesting to note that, in moving from Eqs.(\ref{hartreetransport})
 and (\ref{hartreeconstraint}) to Eqs.(\ref{vlasov}) and (\ref{hatconstr}),
the commutator terms in Eqs.(\ref{hartreetransport}) and
(\ref{hartreeconstraint}) do not contribute.   One sees therefore
explicitly that the
constraint equation (\ref{hartreeconstraint}) has no multiplicative factors
of $\hbar$, while the transport equation contains a simple power of
$\hbar$ throughout.
This is a particular feature of the covariant approach to quantum transport
theories which has been known for some time \cite{vasak,heinzpf}.

The resulting set of equations (\ref{vlasov}) and (\ref{hatconstr}),
as is evident, forms the
{\it classical} limit of the set of  transport and constraint equations in the 
Hartree approximation, since  $m(X)$ is a classical quantity.  
We note that Eq.(\ref{hatconstr}) is trivially fulfilled for the
quasiparticle approximation.


\section{Collisional self-energy}

In this section, we come to the main new aspects  of the paper.   Here
our task is to calculate the collisional self-energy for the model, and to
manipulate the resulting equations in the semiclassical limit into the
form of a Boltzmann-like equation, thereby explicitly exposing the 
weaknesses of the quasiparticle approximation and explicitly showing the
connection between the self-energies evaluated in this approach and the
cross-section that appears in the transport equation.   In particular, we
wish to identify the Feynman graphs that are present in the cross-section,
and their relation to the graphs evaluated for the self-energy.
Note that
some calculations that evaluate the collision term already exist in the
literature \cite{heinz,schoen,perry}.    However, in some cases, in which
the standard spinor decomposition is used \cite{perry}, the connection to
cross-sections is not evident and the resulting expressions are unwieldy and
model-specific.   In others, see for example \cite{schoen}, although it is
somewhat simpler to connect the result to  cross-sections, this has not
been done in as general a  fashion as will be presented here.

To be specific, we investigate  self-energy graphs that 
contain two interaction lines, and which form the contributions to lowest
order to the {\it off-diagonal} matrix elements of $\underline \Sigma$, and
which are illustrated for $\Sigma^{-+}$ in Fig.~4.  The diagram in
Fig.~4(a),
which we denote as the ``direct'' graph, is a term of order $1/N_c$.   The exchange
graph of Fig.~4(b) has the order $1/N_c^2$.    Nevertheless, we retain this
graph, noting that this correction still renders our calculation correct to
$O(1/N_c)$.    As will be seen, the exchange graph plays an essential role.
It leads directly to the cross terms in the scattering amplitudes that we
will encounter.
 We note that
in a rigorous $1/N_c$ expansion, 
 the propagators occurring in Fig.~4 are now
{\it not} to be recalculated self-consistently,
but are to be replaced by the self-consistent Hartree propagators that were
already determined in the previous section. 
In addition,   the simple use of the bare 
interaction, $G$, in Fig.~4 at this level is too naive:  there are further terms
that are also of the same order in $1/N_c$ that must also be included for
completeness \cite{quack,lemmer}. 
   Nevertheless, this simplification, which corresponds to a kind
of Born approximation is a first step in
understanding the role of these diagrams  in a clear fashion,
 and which paves the way for
its generalization to include mesonic intermediate states fully \cite{ski}.
The particular highlight of 
    this 
 study is that 
 illustrates explicitly
how the relativistically formulated transport theory in the semiclassical
limit
results in  a Boltzmann-like collision term, that can be expressed in
terms of a scattering cross section which in turn can be evaluated from a 
well-defined set of Feynman graphs.   We do this 
without introducing the added complexity of the complete set of $1/N_c$ terms
that correspond to the inclusion of mesons.   We have nevertheless alluded
to this extension by our use of notation, labelling the scalar and 
pseudoscalar interaction vertices with a $\sigma$ or $\pi$ respectively.
   The full extension of this
method to include the mesonic degrees of freedom will be handled in a 
separate publication \cite{ski}.

We will now evaluate the first term 
of the collision integral Eq.(\ref{collision}) within  the derivative
expansion with $\hat\Lambda=1$, in order to establish the
order in $\hbar$ of this term.     A further analysis of the
remaining  terms 
that occur in $I_-$ can then be made.

In order to evaluate the collision integral, Eq.(\ref{collision}), we 
require the off-diagonal self-energies.    Consider say $\Sigma^{+-}$.
This can be decomposed into three parts,
\begin{equation}
\Sigma^{+-}(X,p) = \Sigma_\sigma^{+-}(X,p) + \Sigma_\pi^{+-}(X,p) +
 \Sigma^{+-}_{{\rm
mixed}}(X,p),
\label{threesig}
\end{equation}
and similarly for $\Sigma^{-+}$.    Here, $\Sigma_\sigma^{+-}$ and 
$\Sigma_\pi^{+-}$ refer to Fig.~4 in which all vertices have a factor
unity or $i\gamma_5$ in spinor space
attributed to them respectively.    $\Sigma^{+-}_{{\rm mixed}}$
contains the vertices $1$ and $i\gamma_5$ in a mixed fashion and is
shown graphically in Fig.~5.  Note that the direct graphs do not
appear here, as they are identically zero in the quasiparticle approximation.
   Direct translation of the diagrams
of Fig.~4 and 5
in $x$-space can be performed, and then Wigner transformed.  This results in
\begin{eqnarray}
\Sigma_\sigma^{+-}(X,p) &=& - 4G^2\hbar^2\int \frac{d^4p_1}{(2\pi\hbar)^4}
\frac{d^4p_2}{
(2\pi\hbar)^4}\frac{d^4p_3}{(2\pi\hbar)^4} (2\pi\hbar)^4\delta(p-p_1+p_2-p_3) \nonumber \\
      &\times& [ S^{+-}(X,p_1) {\rm tr}\left( S^{-+}(X,p_2) S^{+-}(X,p_3)
\right)
\nonumber \\
& &-S^{+-}(X,p_1)S^{-+}(X,p_2)S^{+-}(X,p_3)  ],
\label{sigsig}
\end{eqnarray} 
while
\begin{eqnarray}
\Sigma_\pi^{+-}(X,p) &=& -4G^2\hbar^2\int\frac{d^4p_1}{(2\pi\hbar)^4}\frac{d^4p_2}{
(2\pi\hbar)^4}\frac{d^4p_3}{(2\pi\hbar)^4} (2\pi\hbar
)^4\delta(p-p_1+p_2-p_3) \nonumber \\
&\times&[\gamma_5 S^{+-}(x,p_1) \gamma_5{\rm tr}\left(
S^{-+}(X,p_2) \gamma_5 S^{+-}(X,p_3) \gamma_5 \right) 
\nonumber \\
& & - \gamma_5 S^{+-}(X,p_1)\gamma_5 S^{-+}(X,p_2)\gamma_5
 S^{+-}(X,p_3) \gamma_5]
\label{sigpi}
\end{eqnarray}
and
\begin{eqnarray}
\Sigma_{{\rm mixed}}^{+-}(X,p) &=& -
4G^2\hbar^2\int\frac{d^4p_1}{(2\pi\hbar)^4}\frac{d^4p_2}{
(2\pi\hbar)^4}\frac{d^4p_3}{(2\pi\hbar)^4} (2\pi\hbar)^4\delta(p-p_1+p_2-p_3) \nonumber \\
&\times& [\gamma_5 S^{+-}(X,p_1) S^{-+}(X,p_2)\gamma_5 S^{+-}(X,p_3) 
\nonumber \\
& & + S^{+-}(X,p_1) \gamma_5 S^{-+}(X,p_2) S^{+-}(X,p_3) \gamma_5].
\label{sigmixed}
\end{eqnarray}
The corresponding expressions for $\Sigma^{-+}$ are obtained from these
by exchanging $+$ and $-$.

Let us now construct  the transport
 equation for $f_q(X,p)$.   As was
indicated in Section \ref{harenergy}, this follows on integrating 
the transport equation for the Green function over the
energy variable $p_0$ on an interval $\Delta_+$ containing $E_p$
and taking the trace.  
Applying this procedure to $I_{{\rm coll}}$, we define
\begin{eqnarray}
J_{{\rm coll}}^q &=& {\rm tr}\int_{\Delta_+} dp_0 I_{{\rm coll}}\nonumber \\
 &=& \int_{\Delta_+} dp_0 {\rm tr}\left(\Sigma^{-+}(X,p)S^{+-}(X,p)\right)
-\int_{\Delta_+} dp_0 {\rm tr}\left(\Sigma^{+-}(X,p)S^{-+}(X,p)\right) 
\nonumber \\
&=& J^{q,{\rm gain}}_{{\rm coll}} - J^{q,{\rm loss}}_{{\rm coll}}
\label{jcoll}
\end{eqnarray}
having set $\hat\Lambda=1$.
Now, on taking cognisance of Eq.(\ref{threesig}), it is clear that each of these terms
itself is to be constructed of three components:
\begin{equation}
J_{{\rm coll}}^{q,{\rm gain/loss}} = J_{{\rm coll},\sigma}^{q,{\rm
gain/loss}} + J_{{\rm coll},\pi}^{q,{\rm
gain/loss}} + J_{{\rm coll,mixed}}^{q,{\rm
gain/loss}}
\label{jsub}
\end{equation}
in which the corresponding subindex is placed on $\Sigma^{-+}$ or
$\Sigma^{+-}$.   In the following section, we shall construct the loss term,
and do this in
detail for the scalar-scalar interaction, $J^{q,{\rm loss}}_{{\rm
coll},\sigma}$ in particular.   The calculations for the remaining terms
then follow by analogy, and these are then sketched 
briefly.

\subsection{The loss term}\label{secloss}

\paragraph{The scalar-scalar interation.}

The scalar-scalar interaction contributes to the loss term  via
\begin{equation}
J^{q,{\rm loss}}_{{\rm coll},\sigma} = \frac{2\pi i}{2E_p}{\rm tr}\left[
\Sigma^{+-}_\sigma(X,p_0=E_p,\vec p) \sum_s u_s(p)\bar u_s(p) f_q(X,\vec p)
\right],
\label{whole}
\end{equation}
on inserting Eq.(\ref{decomp}) into the definition Eq.(\ref{jcoll}), and 
assuming no spin dependence.    One may now insert $\Sigma_\sigma^{+-}
(X,p_0=E_p,\vec p)$ from Eq.(\ref{sigsig}) into Eq.(\ref{whole})
 and multiply out the resulting
expression.   This rather lengthy result contains eight terms,
\begin{eqnarray}
J^{q,{\rm loss}}_{{\rm coll},\sigma} = \frac {\pi i}{E_p}
\frac{4iG^2}{\hbar}
\int\frac{d^4p_1}{(2\pi\hbar)^4}\frac{d^4p_2}{(2\pi\hbar)^4}
\frac{d^4p_3}{(2\pi\hbar)^4} &{}&
(2\pi\hbar)^4
\delta(E_p-p_1^0+p_2^0-p_3^0) \nonumber \\ 
\times \delta(\vec p - \vec p_1 + \vec p_2
-\vec p_3) 
& &(2\pi \hbar)^3
\frac 1{2E_{p_1}}\frac1{2E_{p_2}}\frac 1{2E_{p_3}}\sum_{i=1}^8
T_i
\label{horror}
\end{eqnarray}
with
\begin{eqnarray}
T_1 &=& \delta(p_1^0-E_{p_1})\delta(p_2^0-E_{p_2})\delta(p_3^0-E_{p_3})
[{\rm tr} (u\bar u)_p(u\bar u)_{1}{\rm tr}(u\bar u)_{2}(u\bar u)_{3}
\nonumber \\
& & - {\rm tr} (u\bar u)_p(u\bar u)_{1}(u\bar u)_{2}(u\bar u)_{3}]
\bar f_q(p_1)f_q(p_2)\bar f_q(p_3)f_q(p), \label{t1} \\
T_2 &=& \delta(p_1^0-E_{p_1})\delta(p_2^0-E_{p_2})\delta(p_3^0+E_{p_3})
[{\rm tr} (u\bar u)_p(u\bar u)_{1}{\rm tr}(u\bar u)_{2}(v\bar v )_{-3}
\nonumber \\
& & - {\rm tr} (u\bar u)_p(u\bar u)_{1}(u\bar u)_{2}(v\bar v)_{-3}]
\bar f_q(p_1)f_q(p_2)f_{\bar q}(-p_3)f_q(p), \label{t2} \\
T_3 &=& \delta(p_1^0-E_{p_1})\delta(p_2^0+E_{p_2})\delta(p_3^0-E_{p_3})
[{\rm tr} (u\bar u)_p(u\bar u)_{1}{\rm tr}(v\bar v)_{-2}(u\bar u)_{3}
\nonumber \\
& & - {\rm tr} (u\bar u)_p(u\bar u)_{1}(v\bar v)_{-2}(u\bar u)_{3}]
\bar f_q(p_1)\bar f_{\bar q}(-p_2)\bar f_q(p_3)f_q(p), \label{t3} \\
T_4 &=& \delta(p_1^0-E_{p_1})\delta(p_2^0+E_{p_2})\delta(p_3^0+E_{p_3})
[{\rm tr} (u\bar u)_p(u\bar u)_{1}{\rm tr}(v\bar v)_{-2}(v\bar v)_{-3}
\nonumber \\
& & - {\rm tr} (u\bar u)_p(u\bar u)_{1}(v\bar v)_{-2}(v\bar v)_{-3}]
\bar f_q(p_1)\bar f_{\bar q}(-p_2) f_{\bar q}(-p_3)f_q(p), \label{t4} \\
T_5 &=& \delta(p_1^0+E_{p_1})\delta(p_2^0-E_{p_2})\delta(p_3^0-E_{p_3})
[{\rm tr} (u\bar u)_p(v\bar v)_{-1}{\rm tr}(u\bar u)_{2}(u\bar u)_{3}
\nonumber \\
& & - {\rm tr}(u\bar u)_p(v\bar v)_{-1}(u\bar u)_{2}(u\bar u)_{3}]
f_{\bar q}(-p_1)f_q(p_2)\bar f_q(p_3)f_q(p), \label{t5} \\
T_6 &=& \delta(p_1^0+E_{p_1})\delta(p_2^0-E_{p_2})\delta(p_3^0+E_{p_3})         
[{\rm tr} (u\bar u)_p(v\bar v)_{-1}{\rm tr}(u\bar u)_{2}(v\bar v)_{-3}
\nonumber \\
& & - {\rm tr}(u\bar u)_p(v\bar v)_{-1}(u\bar u)_{2}(v\bar v)_{-3}]
f_{\bar q}(-p_1)f_q(p_2) f_{\bar q}(-p_3)f_q(p), \label{t6} \\
T_7 &=& \delta(p_1^0+E_{p_1})\delta(p_2^0+E_{p_2})\delta(p_3^0-E_{p_3})         
[{\rm tr} (u\bar u)_p(v\bar v)_{-1}{\rm tr}(v\bar v)_{-2}(u\bar u)_{3}
\nonumber \\
& & - {\rm tr}(u\bar u)_p(v\bar v)_{-1}(v\bar v)_{-2}(u\bar u)_{3}]
f_{\bar q}(-p_1)\bar f_{\bar q}(-p_2)\bar f_q(p_3)f_q(p), \label{t7} \\
T_8 &=& \delta(p_1^0+E_{p_1})\delta(p_2^0+E_{p_2})\delta(p_3^0+E_{p_3})          
[{\rm tr} (u\bar u)_p(v\bar v)_{-1}{\rm tr}(v\bar v)_{-2}(v\bar v)_{-3}
\nonumber \\                                             
& & - {\rm tr}(u\bar u)_p(v\bar v)_{-1}(v\bar v)_{-2}(v\bar v)_{-3}]
f_{\bar q}(-p_1)\bar f_{\bar q}(-p_2) f_{\bar q}(-p_3)f_q(p)], \label{t8} 
\end{eqnarray}
where spin indices have been suppressed.  Here, the indices ``$i$'' mean
$p_i$, $i=1,3$, while the notation ``$-i$'' refer to the negative 
four-momenta, $-p_i$.
At first sight, this expression appears formidable.   It provides us, 
however, with some physical insight.   Each term contains a product of a 
combination of four quark and antiquark distribution functions, with
$f_q(p)$ being present in all of them, being the external function under
study.    One may attribute a diagram to each of these processes in a loose
sense, by assigning incoming quark/antiquark lines to $f_{q,\bar q}$ and
outgoing ones to $\bar f_{q,\bar q}$.     In doing so, it appears that 
quark-quark scattering terms arise from $T_1$, and that quark-antiquark
scattering graphs arise from $T_4$ and $T_7$.    The remaining graphs
correspond to the processes  shown in Fig.~6, and  which involve
particle and antiparticle creation / annihilation.  Note that time runs
from  left to right in these diagrams.
We emphasise that the graphs depicted in this figure are {\it not}
Feynman graphs.   The identification of Feynman graphs requires a detailed
analysis of the coefficient factors, such as that which will be done for
$T_1$, $T_4$ and $T_7$. 
In Fig.~6, graph (a) corresponds to the process of $T_3$, while (b)
corresponds to those of $T_2$ and $T_5$.    The expression for $T_8$ is
given by (c), while the vacuum fluctuations of (d) correspond to $T_6$.
   Note that these all are categorized as off-shell
processes:   on performing the energy integrations over $dp_1^0$, $dp_2^0$
and $dp_3^0$, it becomes manifest that these terms vanish due to to
energy-momentum conservation.    Thus, in contrast to the non-relativistic
case, one notes that the quasiparticle assumption serves to suppress 
particle / antiparticle creation / annihilation, a restriction that may be 
too severe for application to a fundamental theory of quarks and gluons.

It now remains for us to validate that $T_1$, $T_4$ and $T_7$ indeed
correspond to the processes of quark-quark and quark-antiquark scattering
and to identify the scattering graphs in a strict sense.   Here we wish
to identify actual Feynman diagrams, and not
simply associate a heuristic diagram to the processes occurring in these
terms.    To this end, we require a detailed analysis of the coefficients.
We thus proceed by evaluating the energy integrals, and make the substitution 
$p_i\rightarrow -p_i$ for the antiquark state, to find
\begin{eqnarray}
J_{{\rm coll},\sigma}^{q,{\rm loss}} &=& -\frac{\pi i}{E_p} \frac{4G^2
i^3}\hbar
\int\frac{d^3p_1}{(2\pi\hbar)^3 2E_{p_1}} \frac{d^3p_2}{(2\pi\hbar)^3 2E_{p_2}}
\frac{d^3p_3}{(2\pi\hbar)^3 2E_{p_3}} (2\pi\hbar)^4 \nonumber \\
&\times& [\delta^4(p-p_1+p_2-p_3) ({\rm tr}(u\bar u)_p(u\bar u)_1{\rm tr} (u\bar u)_2
(u\bar u)_3 \nonumber \\
& & \quad - {\rm tr} (u\bar u)_p(u\bar u)_1(u\bar u)_2 (u\bar u_3))
\bar f_q(p_1) f_q(p_2)\bar f_q(p_3) f_q(p) 
\nonumber \\
&+& \delta^4(p-p_1-p_2 +p_3)({\rm tr}(u\bar u)_p(u\bar u)_1 {\rm tr} (v\bar v)_2
(v\bar v)_3 \nonumber \\
& & \quad - {\rm tr}(u\bar u)_p(u\bar u)_1 (v\bar v)_2(v\bar v)_3)
\bar f_q(p_1)\bar f_{\bar q}(p_2)f_{\bar q}(p_3)f_q(p) 
\nonumber \\
&+& \delta^4(p+p_1-p_2-p_3)({\rm tr}(u\bar u)_p(v\bar v)_1{\rm tr} (v\bar
v)_2
(u\bar u)_3 \nonumber \\
& & \quad  - {\rm tr}(u\bar u)_p(v\bar v)_1(v\bar v)_2(u\bar u)_3)
f_{\bar q}(p_1)\bar f_{\bar q}(p_2)\bar f_{q}(p_3) f_q(p)].
\nonumber
\\
\label{monster}
\end{eqnarray}
This can be simplified slightly, if one makes the substitution
$p_1\leftrightarrow p_3$ in the last term, enabling one to combine the
second and third terms.   At the same time, one notes that there is a 
symnmetry in the first term in the variables $p_1$ and $p_3$.   We employ
this to rewrite the first term also.    One then has the form
\begin{eqnarray}
J_{{\rm coll},\sigma}^{q,{\rm loss}} &=& -\frac{\pi i}{E_p} \frac{4G^2
i^3}\hbar
\int\frac{d^3p_1}{(2\pi\hbar)^3 2E_{p_1}} \frac{d^3p_2}{(2\pi\hbar)^3 2E_{p_2}}
\frac{d^3p_3}{(2\pi\hbar)^3 2E_{p_3}} (2\pi\hbar)^4 \nonumber \\
&\times&\{\delta^4(p-p_1+p_2-p_3)\frac 12[{\rm tr} (u\bar u)_p(u\bar u)_1
{\rm tr}(u\bar u)_2(u\bar u)_3  \nonumber \\
& & \quad + {\rm tr}(u\bar u)_p(u\bar u)_3{\rm tr}(u\bar u
)_2(u\bar u)_1 - {\rm tr} (u\bar u)_p(u\bar u)_1(u\bar u)_2(u\bar u)_3
\nonumber \\ 
& &\quad  
 - {\rm tr}
(u\bar u)_p(u\bar u)_3(u\bar u)_2 (u\bar u)_1] 
 \bar f_q(p_1) f_q(p_2)\bar f_q(p_3) f_q(p)
\nonumber \\
&+&\delta^3(p-p_1-p_2+p_3) [{\rm tr}(u\bar u)_p(u\bar u)_1{\rm tr}
(v\bar v)_2(v\bar v)_3 \nonumber \\
& & \quad + {\rm tr}(u\bar u)_p(v\bar v)_3{\rm tr}(v\bar v)_2
(u\bar u)_1 - {\rm tr}(u\bar u)_p(u\bar u)_1(v\bar v)_2(v\bar v)_3 
\nonumber \\
& &\quad - {\rm tr}
(u\bar u)_p(v\bar v)_3(v\bar v)_2(u\bar u)_1]
\bar f_q(p_1)\bar f_{\bar q}(p_2) f_{\bar q}(p_3)f_q(p)\}
\nonumber \\
\label{fhorror}
\end{eqnarray}
Now, by directly examining the quark-quark scattering processes, see Fig.~7,
 one sees that the first two terms that arose from the {\it direct} graph of
Fig.~4(a)
 correspond to the $t$ and $u$ channel
exchange diagrams respectively, while the remaining terms
that arose from the exchange graph of Fig.~4(b), are constructed  from the 
product combinations of the $t$ and $u$ channel matrix elements.
The individual matrix elements for these processes are
\begin{eqnarray}
-iM_\sigma^t(qq\rightarrow qq) &=& \frac{2iG}\hbar
\bar u_1^\alpha u_p^\alpha\bar u_3^\beta
u_2^\beta \label{mqq1} \\
-iM_\sigma^u(qq\rightarrow qq) &=& \frac{2iG}\hbar 
\bar u_1^\alpha u_2^\alpha\bar u_3^\beta
u_p^\beta .
\label{mqq}
\end{eqnarray}
    Similarly, in the quark antiquark channel shown in Fig.~8,
the $s$ and $t$ channel graphs are represented together with their
 products.   Their
individual matrix elements are
\begin{eqnarray}
-iM_\sigma^s(q\bar q\rightarrow q\bar q) &=& \frac{2iG}\hbar\bar u_1^\alpha v_2^\alpha\bar
v_3^\beta u_p^\beta 
\\
-iM_\sigma^t(q\bar q\rightarrow q\bar q) &=& \frac{2iG}\hbar
\bar u_1^\alpha u_p^\alpha \bar
v_3^\beta v_2^\beta .
\label{mqqb}
\end{eqnarray}
Squaring these matrix elements and
comparing the result with the terms in Eq.(\ref{fhorror}), one arrives at the
simple form
\begin{eqnarray}
J_q^{{\rm loss},\sigma} = &-& \frac{\pi\hbar}{E_p} \int\frac{d^3p_1}{(2\pi\hbar)^3
2E_{p_1}} \frac{d^3p_2}{(2\pi\hbar)^3 2E_{p_2}}
\frac{d^3p_3}{(2\pi\hbar)^3 2E_{p_3}} (2\pi\hbar)^4  \nonumber \\
&\times&\{\delta^4(p-p_1+p_2-p_3)\frac 12\sum |M^\sigma_{qq\rightarrow qq}
(p2\rightarrow 13)|^2\bar f_q(p_1)f_q(p_2)\bar f_q(p_3)f_q(p) \nonumber \\
&+&\delta^4(p-p_1-p_2+p_3)\sum |M^\sigma_{q\bar q\rightarrow q\bar q}
(p3
\rightarrow 12)|^2\bar f_q(p_1)\bar f_{\bar q}(p_2)f_{\bar q}(p_3)f_q(p)\},
\nonumber \\
\label{matr}
\end{eqnarray}
where $|M^\sigma_{qq\rightarrow qq}| = |M^t(qq\rightarrow qq) - 
M^u(qq\rightarrow qq)|$ and $|M^\sigma_{q\bar q\rightarrow q\bar q}|
= |M^s(q\bar q\rightarrow q\bar q) - M^t(q\bar q\rightarrow q\bar q)|$
contains the relevant $s$, $t$ and $u$ channels, and 
 the sum refers to all spin variables.

In this calculation, one has explicitly observed that in the quark-quark
scattering, the direct graph of Fig.~4(a) leads to the squares of the 
matrix elements of both the $t$ and $u$ channels, 
while cross terms are attributed to the exchange graph.   Similarly, for the
quark-antiquark processes, the direct graph of Fig.~4(a) leads to the
squares of the matrix elements for the $s$ and $t$ channels, again with
the cross terms arising from Fig.~4(b).   One thus sees the necessity 
of retaining both of these graphs in order to derive the sum of
different channels in the cross section.

\paragraph{The pseudoscalar-pseudoscalar interaction.}   The interaction
graphs of Fig.~4 that contain only the pseudoscalar $i\gamma_5$ at each
vertex contribute to the collision term via
\begin{equation}
J^{q,{\rm loss}}_{{\rm coll},\pi} = \frac {2\pi i}{2E_p} {\rm tr}\left[ 
\Sigma^{+-}_\pi(X,p_0=E_p,\vec p)\sum_s u_s(p)\bar u_s(p) f_q(X,\vec p)
\right],
\label{picoll}
\end{equation}
with $\Sigma_\pi^{-+}$ given in Eq.(\ref{sigpi}).   This term may be dealt
with in 
the same fashion as the scalar term.
Explicit details are left to the  Appendix D.   After some manipulation, 
one finds
\begin{eqnarray}
J_q^{{\rm loss},\pi} = &-& \frac{\pi\hbar}{E_p} \int\frac{d^3p_1}{(2\pi\hbar)^3
2E_{p_1}} \frac{d^3p_2}{(2\pi\hbar)^3 2E_{p_2}}
\frac{d^3p_3}{(2\pi\hbar)^3 2E_{p_3}} (2\pi\hbar)^4  \nonumber \\
&\times&\{\delta^4(p-p_1+p_2-p_3)\frac 12\sum |M^\pi_{qq\rightarrow qq}
(p2\rightarrow 13)|^2\bar f_q(p_1)f_q(p_2)\bar f_q(p_3)f_q(p) \nonumber \\
&+&\delta^4(p-p_1-p_2+p_3)\sum |M^\pi_{q\bar q\rightarrow q\bar q}
(p3
\rightarrow 12)|^2\bar f_q(p_1)\bar f_{\bar q}(p_2)f_{\bar q}(p_3)f_q(p)\},
\nonumber \\
\label{matrpi2}
\end{eqnarray}
where $|M^\pi_{qq\rightarrow qq}| = |M_\pi^t(qq\rightarrow qq) -
M_\pi^u(qq\rightarrow qq)|$ and $|M^\pi_{q\bar q\rightarrow q\bar q}|
= |M_\pi^s(q\bar q\rightarrow q\bar q) 
- M_\pi^t(q\bar q\rightarrow q\bar q)|$
contain  the relevant $s$, $t$ and $u$ channel scattering processes that are
given in Figs.~7 and 8 with the vertices $i\gamma_5$ everywhere.

One can now combine Eqs.(\ref{matr}) and (\ref{matrpi2}).  
These are added. In addition, it is useful to   
 relabel the variables $2\leftrightarrow 3$ in the
quark-antiquark scattering terms, so that one may extract a common
$\delta$-function.    One then has the combined terms
\begin{eqnarray}
J_{{\rm coll},\sigma + \pi}^{{\rm loss},q} &=& J^{q,{\rm loss}}_{{\rm coll},
\sigma} + J_{{\rm coll},\pi}^{q,{\rm loss}} \nonumber \\
&=& -\frac{\pi\hbar}{E_p}\int\frac{d^3p_1}{(2\pi\hbar)^3 2E_{p_1}}
 \frac {d^3p_2}
{(2\pi\hbar)^3 2E_{p_2}}\frac{d^3p_3}{(2\pi\hbar)^3 2E_{p_3}} (2\pi\hbar)^4
 \delta^4(p
-p_1 + p_2 - p_3) \nonumber \\
&\times& \{\frac 12(\sum|M^\sigma_{qq\rightarrow qq}(p2\rightarrow 13)|^2
+ \sum|M_{qq\rightarrow q q}^\pi(p2\rightarrow 13)|^2) \nonumber \\
& & \quad \times \bar f_q(p_1)f_q(p_2)\bar f_q(p_3) f_q(p) \nonumber \\
&+&(\sum|M^\sigma_{q\bar q\rightarrow q\bar q}(p2\rightarrow 13)|^2 + \sum
|M^\pi_{q\bar q\rightarrow q\bar q}(p2\rightarrow 13)|^2) \nonumber \\
& & \quad \times \bar f_q(p_1)\bar f_{\bar q}(p_3) f_{\bar q}(p_2) f_q(p)
\}.
\label{sigmapi}
\end{eqnarray}
From this form, it is evident that both scalar and pseudoscalar 
interaction terms combine to give the individual matrix elements squared.
What is still missing, however, is the mixed self-energy.    This is
examined briefly in the following section.

\paragraph{Mixed interaction.}  The mixed self-energy, given in
Eq.(\ref{sigmixed}) contributes to the loss term via
\begin{equation}
J^{q,{\rm loss}}_{{\rm coll, mixed}} = \frac{2\pi i}{2E_p}{\rm tr}\left[
\Sigma_{{\rm mixed}}^{+-}(X,p_0=E_p,\vec p)\sum_s u_s(p)\bar u_s(p) f_q(X,
\vec p)
\right].
\label{lossmix}
\end{equation}
Dealing with this term as with the scalar interaction, one finds a series
of terms, the details of which are to be found in Appendix E.   One can
identify these in terms of the amplitudes of the 
scalar and pseudoscalar scattering matrix elements, see Eqs.(\ref{mqq}),
(\ref{mqqb}), (\ref{mqq2}) and (\ref{mqq3}) 
and one finds
\begin{eqnarray}
J^{q,{\rm loss}}_{{\rm coll, mixed}} &=& -\frac {\pi\hbar}
{E_p}\int \frac{d^3p_1}
{(2\pi\hbar)^3 2E_{p_1}}\frac{d^3p_2}{(2\pi\hbar)^3 2E_{p_2}}
\frac{ d^3p_3}{(2\pi\hbar)^3
2E_{p_3}} (2\pi\hbar)^4 \delta(p-p_1+p_2-p_3) \nonumber \\
&\times&\{\frac12 [\sum M_\sigma^u M_\pi^t{}^* + \sum M_\sigma^t M_\pi^u{}^*
+\sum M_\pi^t M_\sigma^u{}^* + \sum M_\pi^u M_\sigma^t{}^*]_{qq\rightarrow
qq} \nonumber \\
& & \quad\quad \times \bar f_q(p_1) f_q(p_2)\bar f_q(p_3) f_q(p) \nonumber\\
&+& [\sum M_\sigma^s M_\pi^t{}^* + \sum M_\sigma^t M_\pi^s{}^* + \sum M_\pi^t
M_\sigma^s{}^* + \sum M_\pi^s M_\sigma^t{}^*]_{q\bar q\rightarrow q\bar q} 
\nonumber \\
& & \quad\quad \times \bar f_q(p_1) \bar f_{\bar q}(p_3) f_{\bar q} (p_2) f
_q(p) \}. \nonumber \\
\label{wonder}
\end{eqnarray}
Thus, the final result for the collision term,
$J^{q,{\rm loss}}_{{\rm coll}}$, constructed by adding Eq.(\ref{wonder})
to Eq.(\ref{sigmapi}), is
\begin{eqnarray}
J^{q,{\rm loss}}_{{\rm coll}} &=& -\frac {\pi\hbar} {E_p} 
\int\frac{d^3p_1}{(2\pi\hbar)^3 2E_{p_1}} \frac {d^3p_2}
{(2\pi\hbar)^3 2E_{p_2}}\frac{d^3p_3}{(2\pi\hbar)^3 2E_{p_3}} (2\pi\hbar)^4  
\delta^4(p-p_1+p_2 -p_3) \nonumber \\
& & \times\{\frac 12\sum |M_{qq\rightarrow qq}(p2\rightarrow 13)|^2\bar f_q(p_1)
f_q(p_2)\bar f_q(p_3) f_q(p) \nonumber \\
& & \quad + \sum |M_{q\bar q\rightarrow q\bar q}(p2\rightarrow 13)|^2 \bar f
_q(p_1)\bar f_{\bar q}(p_3)f_{\bar q}(p_2) f_q(p) \} ,
\label{final}
\end{eqnarray}
where the 
matrix elements are now complete in both the quark-quark and quark-antiquark
sectors,
 $M_{qq\rightarrow qq} = M_{qq\rightarrow qq}^\sigma + 
M_{qq\rightarrow qq}^\pi$ and $M_{q\bar q\rightarrow q\bar q} =
M^\sigma_{q\bar q\rightarrow q\bar q} + M^\pi_{q\bar q\rightarrow
q\bar q}$.    We comment once again that although this result has
been derived using a constant interaction or in Born approximation,
it will also hold in an
 extended form, where the interaction proceeds rather via
a momentum dependent coherent mesonic state
 \cite{ski}.

\subsection{The gain  term} \label{gainterm}

Construction of the gain term proceeds exactly as in the loss case
with the evaluation of 
\begin{equation}
J^{q,{\rm gain}}_{{\rm coll}} = -\frac {2\pi i}{2E_p}
{\rm tr} \left[\Sigma ^{-+}(X,p_0=E_p,\vec p)\sum_s u_s(p) \bar u_s(p)
\bar f_q(X,\vec p)\right]
\label{gain}
\end{equation}
for the three different forms of $\Sigma^{-+}$.    We do not repeat this 
procedure here, but simply quote the final result.    One has
\begin{eqnarray}
J^{q,{\rm gain}}_{{\rm coll}} &=& -\frac {\pi\hbar} {E_p}
\int\frac{d^3p_1}{(2\pi\hbar)^3 2E_{p_1}} \frac {d^3p_2}
{(2\pi\hbar)^3 2E_{p_2}}\frac{d^3p_3}{(2\pi\hbar)^3 2E_{p_3}} (2\pi\hbar)^4 
\delta^4(p-p_1+p_2 -p_3) \nonumber \\
& & \times\{\frac 12\sum |M_{qq\rightarrow qq}(p2\rightarrow 13)|^2 f_q(p_1)
\bar f_q(p_2) f_q(p_3)\bar f_q(p) \nonumber \\
& & \quad  + \sum |M_{q\bar q\rightarrow q\bar q}(p2\rightarrow 13)|^2  f
_q(p_1) f_{\bar q}(p_3)\bar f_{\bar q}(p_2) \bar f_q(p) \}
\label{finalg}
\end{eqnarray} 
with the complete quark-quark and quark-antiquark scattering
matrix elements as defined in Eq.(\ref{final}).    

\subsection{The transport equation}

In this section, our aim is to construct the transport equation that now
includes the self-energies represented in Figs.4 and 5.    In order to do
so, we first combine the results of Section \ref{secloss} and 
\ref{gainterm} to construct the collision integral 
 $J_{{\rm coll}}^q$ that was defined by Eq.(\ref{jcoll}).
This energy integrated collision term can then be written as
\begin{eqnarray}
&J&_{{\rm coll}}^q = -\frac{\pi\hbar}{E_p}\int dQ \int
\frac{d^3p_2}{(2\pi\hbar)^32E_{p_2}}
 \nonumber \\
&\times&\{ \frac 12\sum|M_{qq\rightarrow qq}(p2\rightarrow 13)|^2 
(f_q(p_1) \bar f_q(p_2) f_q(p_3)\bar f_q(p) - \bar f_q(p_1) f_q(p_2)\bar f
_q(p_3) f_q(p)) \nonumber \\
&+& \sum|M_{q\bar q\rightarrow q \bar q}(p2\rightarrow 13)|^2(
f_q(p_1)\bar f_{\bar q}(p_2) f_{\bar q}(p_3)
\bar f_q(p) -
\bar f_q(p_1)
\bar f_{\bar q}(p_3) f_{\bar q}(p_2) f_q(p) \},
\nonumber \\
\label{collterm}
\end{eqnarray}
where
\begin{equation}
dQ = (2\pi\hbar)^4\delta^4(p+p_2-p_1-p_3)\frac{d^3p_1}{(2\pi\hbar)^3 2E_{p_1}}
\frac{d^3p_3}{(2\pi\hbar)^3 2E_{p_3}}
\end{equation}
is the Lorentz invariant phase space factor.
   One can, however, reexpress the right hand side of this
equation in terms of the differential scattering cross section.  One can
see this simply if one considers centre of mass variables: noting that
\begin{equation}
dQ = \frac 1{4\pi^2\hbar^2}\frac{|p_1|}{4\sqrt s} d\Omega
\label{dq}
\end{equation}
and
\begin{equation}
\frac{d\sigma}{d\Omega}  = \frac 1{64\pi^2\hbar^2 s}\frac{|p_1|}{|p_2|} |\bar M|^2,
\label{dsigma}
\end{equation}
where $|\bar M|^2$ is the {\it spin and color averaged } scattering matrix
element squared.
    One thus has
\begin{eqnarray}
J^q_{{\rm coll}} &=& -\frac{4N_c^2\pi\hbar}{E_p} \int d\Omega  
\int \frac {d^3p_2}{(2\pi\hbar)^3 2E_{p_2}} |\vec v_p-\vec v_2|
2E_p 2E_{p_2}
\nonumber \\
&\times&\{ \frac 12
\frac{d\sigma}{d\Omega}|_{qq\rightarrow qq}(p2\rightarrow 13)[f_q(p_1)
\bar f_q(p_2)
f_q
(p_3) \bar f_q(p) - \bar f(p_1) f_q(p_2) \bar f_q(p_3) f_q(p)] \nonumber \\
& & +\frac{d\sigma}{d\Omega}|_{q\bar q\rightarrow q\bar q}(p2\rightarrow 13)
[f_q(p_1) \bar
f_{\bar q}(p_2) f_{\bar q}(p_3) \bar f_q(p) -
\bar 
f_q(p_1) \bar f_{\bar q}(p_3) f_{\bar q}(p_2) f_q(p) 
]\} \nonumber \\
\label{cof}
\end{eqnarray}
for the collision term.   We have thus obtained the final expression for 
the collision integral that looks formally similar to that expected in the
semi-classical limit.   In addition, we note that this term is proportional
to $\hbar$, explicitly demonstrating our {\it a priori} claim of
Sec.~III~C that terms containing off-diagonal matrix elements of $\Sigma$ 
multiplied with $S$ are proportional to $\hbar$.

In view of this, we may examine the right hand side of the transport 
equation
 Eq.(\ref{full1}) that contains
 $I_-$ of Eq.(\ref{is}).     We require the energy integrated and traced
form of $I_-$, i.e. 
\begin{equation}
J_-^q = J_{{\rm coll}} + J_-^{q,A} + J^{q,R}_-,
\label{js}
\end{equation}
with
\begin{equation}
J_-^{q,A/R} = {\rm tr} \int_{\Delta_+} dp_0 I_-^{A,R}.
\label{defj}
\end{equation}
We are required to consider
$J_-^{q,A}$ and $J_-^{q,R}$ (or alternatively $I_-^R$ and $I_-^A$) also
 {\it only
to this first  order} of $\hbar$ for this term in the self-energy.    Examining
Eqs.(\ref{irmp}) and (\ref{iamp}), one notes that $J_-^{q,A}$ and
$J_-^{q,R}$ then vanish identically, due to the cyclic property of the
trace.
Higher orders would be of order $O(\hbar^2)$, and may therefore be
discarded.
Thus the transport equation Eq.(\ref{hartreetransport}) now becomes modified to
read  
\begin{eqnarray}
\frac {i\hbar}2\{\gamma^\mu,\frac{\partial S^{-+}(X,p)}{\partial X^\mu}\}
+[\not p,S^{-+}(X,p)] =  - i\hbar \partial_\mu \Sigma_H^A(X)
\partial_p^\mu S^{-+}(X,p) + I_{{\rm coll}},
\label{transport}
\end{eqnarray}
with $I_{{\rm coll}}$ related to $J_{{\rm coll}}$ via Eq.(\ref{defj}) and
$J_{{\rm coll}}$ as given in Eq.(\ref{cof}). 
It becomes clear, on integrating over the
positive energies $p_0$ and performing the spinor trace,
\begin{equation}
\frac{i\hbar}2\int_{\Delta_+} dp_0\  {\rm tr} \{\gamma^\mu,\frac{\partial
S^{-+}(X,p)}{\partial X^\mu}\}=  - i\hbar  \int_{\Delta_+} dp_0 \ 
\partial_\mu \Sigma_H^A(X)
\partial_p^\mu S^{-+}(X,p) + J_{{\rm coll}}
\label{whew}
\end{equation}
that all terms that survive this procedure,
 are directly proportional to $\hbar$, and that
the resulting transport equation
\begin{eqnarray}
p^\mu\partial_\mu f_q&(&X,\vec p) + m(X)\partial_\mu m(X)\partial_p^\mu
f_q(X,
\vec p) =
N_c\int d\Omega \int \frac {d^3p_2}{(2\pi\hbar)^3 2E_{p_2}} |\vec v_p-\vec v_2| 
2E_p 2E_{p_2}
\nonumber \\
&\times&\{ \frac 12
\frac{d\sigma}{d\Omega}|_{qq\rightarrow qq}(p2\rightarrow 13)(f_q(p_1)\bar f_q(p_2) f_q
(p_3) \bar f_q(p) - \bar f(p_1) f_q(p_2) \bar f_q(p_3) f_q(p)) \nonumber \\
& & +\frac{d\sigma}{d\Omega}|_{q\bar q\rightarrow q\bar q}(p2\rightarrow 13)
(f_q(p_1) \bar
f_{\bar q}(p_2) f_{\bar q}(p_3) \bar f_q(p)
- \bar
f_q(p_1) \bar f_{\bar q}(p_3) f_{\bar q}(p_2) f_q(p))\}, \nonumber \\
\label{boltmann}
\end{eqnarray}
is the classical result.   Here this equation has been
cast into a form that is reminiscent of the non-relativistic
Boltzmann equation \cite{reif}.  To make this analogy exact, one should 
regard this equation rather in terms of the quark density distribution, that is
given as $n_q(p,X) = 2N_cf_q(p,X)$.
   It is thus apparent that the lowest
order term in the derivative expansion of $I_{{\rm coll}}$
 gives the relativistic analogue of
the classically expected collision term \cite{reif}.    By contrast,
the first order term was required in the local self-energy in deriving the
force term from the Hartree approximation. 
   Finally, we comment that the transport equation
for the antiquark distribution function
 is simply obtained from the aforegoing by replacing
$q\leftrightarrow\bar q $ everywhere.

Before concluding this section, we make a comment with regard to the
relevance of chiral symmetry to Eq.(\ref{boltmann}).    Formally, any
arbitrary interaction will lead to a similar form as has been given in
Eq.(\ref{boltmann}) for the transport equation, and therein lies the
generality of this calculation.   However, a different self-energy would
describe 
the mean field part or Vlasov
part.  For the NJL model, this amounts to a 
 dynamically generated quark mass $m(X)$ appearing, that arises because of
 the symmetry breaking. 
     Note that in the Wigner-Weyl realization of the
symmetry, this term would vanish identically.   In addition, the issue of
chiral symmetry reflects itself again in the cross-sections that are to
be calculated for Eq.(\ref{boltmann}), see Ref.\cite{scatt}.

\subsection{The constraint equation}

The constraint equation, Eq.(\ref{full2}), can also be examined in light of
the $\hbar$ expansion.   One notes that it is {\it intrinsically of one
order less than} the transport equation in terms of an $\hbar$ expansion:
the semiclassical form of Eq.(\ref{full2}) arises from $\hbar^0$ structures.
Thus, in extending the Hartree equation to include collisional
self-energies, one may immediately disregard the collisional term $I_{{\rm
coll}}$, as well as 
$I^R_+$ since they are one order higher in $\hbar$ than the Hartree
approximation.    An additional correction to
the mass could at most arise from $I^A_+$ but which represents $O(1/N_c)$
corrections to $m(X)=m_{\rm H}(X)$ calculated via Eq.(\ref{gap}).
   In the absence
of this term, the constraint equation retains the form already given in
Eq.(\ref{hatconstr})

\section{Summary and Conclusions}

In this paper, we have presented a general derivation for transport and
constraint equations for a fermionic system from a canonical point of view
along the lines of Ref.\cite{heinz}.
We have evaluated these for the single flavor NJL model.   Here two 
expansion schemes are overlaid.     The strong coupling theory is dealt with
via an expansion in the inverse number of colors, $1/N_c$, in what is by now
standard practice.   We make an expansion in $\hbar$, whch only 
partly is identical with the gradient expansion.
  We have evaluated the
lowest order self-energy term in the $1/N_c$ expansion, and constructed
transport (Vlasov) and constraint (mass-shell) equations.   In addition,
that one is working with a chiral model evidences itself in that 
 the gap equation must be solved in addition
to the transport and constraint equation emerges, and must be solved 
together with these.
It is demonstrated  explicitly that differing orders in $\hbar$ are required
for these two equations
\footnote{Although this
feature is known in the literature \cite{vasak}, it is not often made
explicit \cite{heinzpf}.},
 the constraint equation being of order $\hbar^0$
while the transport equation has order $\hbar^1$.

We then examine a term of the next order in the $1/N_c$ expansion.
We treat the simplest one that gives rise to a non-vanishing collision 
integral, and show explicitly that it is of order $\hbar$, and therefore
is necessary for constructing a consistent transport equation.   By the same
token, it has no relevance for the constraint equation, for which only 
additional mass changes can occur.

In the derivation of the relativistic Boltzmann equation, we have gone to
some trouble to indicate explicitly that the on-mass shell condition
 restricts the collision integral to contain only elastic
scattering processes, and suppresses pair creation and annihilation
amplitudes $q\leftrightarrow (q\bar q)q$
 and also vacuum fluctuations that would otherwise be present.
The derivation of the collision integral that we have given here 
 is general for all relativistic
two body interactions -- minimal additional manipulations are required for
a momentum dependent interaction.    Once again, the addition feature of
the chiral symmetry manifests itself in that one has to solve a gap
equation concomitantly.   In addition, the cross-sections are a function of
the dynamically generated quark mass, and therefore also reflect the chiral
symmetry aspect of the problem.

In our calculation, we have also been able to identify the
associated $t$ and $u$ channel matrix elements required to this level of
approximation for quark-quark scattering, as well as the $s$ and $t$ channel
amplitudes for quark-antiquark scattering, i.e. we are able to explicitly
attribute actual Feynman graphs, and not simply heuristic diagrams, to
each of these processes.

In this paper, we have not made any attempt to include the Fock term,
which is also of order $O(1/N_c)$.
This is deliberate.   Since the next stage in this study involves the 
incorporation of the mesonic sector, of which the Fock term forms a part,
it has no relevance here at this stage and can at most modify
the quasiparticle mass arising from  the Hartree
term.   
It is important to include {\it all} diagrams to $O(1/N_c)$, thereby 
incorporating the mesonic sector completely.    This in itself is a complex
task, and contains several other difficulties.
This will be the subject of a future paper \cite{ski}.

\section*{ACKNOWLEDGMENTS}

Two of the authors, S.P.K. and J.H., gratefully acknowledge the various
collaborations with R.H. Lemmer, during which they have profited immensely
from his experience in the techniques of many-body theory.    This paper is
therefore dedicated to him.
This work has been supported in part by the Deutsche Forschungsgemeinschaft
DFG under the contract number Hu 233/4-4, and by the German Ministry for
Education and Research (BMBF) under contract number 06 HD 742.

\appendix

\section{Green functions}

In this section, we list relations among the various Green functions and
the self-energies.   They can, in almost all cases, be simply derived from
their definitions.

The causal and anticausal Green functions, see Eq.(\ref{gfunctions}),
 are related to 
$S^{-+}$ and $S^{+-}$ via the relations
  \begin{equation}                                    
     S^{--}_{\alpha \beta} ( x,y )
         = \theta(x_{0}-y_{0}) S^{+-}_{\alpha \beta} ( x,y ) +
             \theta(y_{0}-x_{0}) S^{-+}_{\alpha \beta} ( x,y ) ,
\label{a1}
  \end{equation}
and
\begin{equation}
   S^{++}_{\alpha \beta} ( x,y )
         = \theta(y_{0}-x_{0}) S^{+-}_{\alpha \beta} ( x,y ) +
             \theta(x_{0}-y_{0}) S^{-+}_{\alpha \beta} ( x,y ) .
\label{a2}
 \end{equation}
The Green functions are not independent.   They are linearly related in a
way
that is obvious from their definitions,
\begin{eqnarray}
S^{--}_{\alpha\beta} (x,y) + S^{++}_{\alpha\beta}(x,y) &=& 
S^{-+}_{\alpha\beta} (x,y) + S^{+-}_{\alpha\beta}(x,y) \nonumber \\
 &=&  S^{\rm K}_{\alpha \beta} ( x,y )
           =
      -\frac i\hbar     \langle \left[ \psi_{\alpha}(x),
                                \overline{\psi}_{\beta}(y) \right] \rangle
\label{a3}
\end{eqnarray}
and which defines the Keldysh Green function.
On the other hand, the retarded and advanced Green functions that were 
introduced in Eqs.(\ref{retard}) and (\ref{advance}) can also be related 
to the $S^{ij}$, $i,j=\pm$, via
  \begin{eqnarray}
     S^{\rm R}_{\alpha \beta} ( x,y )
              &=& \theta(x_{0}-y_{0}) \left\{ S^{+-}_{\alpha \beta} ( x,y ) -
                                     S^{-+}_{\alpha \beta} ( x,y ) \right\}
                                                      \nonumber \\
              &=& S^{--}_{\alpha \beta} ( x,y ) - S^{-+}_{\alpha \beta} (
x,y )
               =  S^{+-}_{\alpha \beta} ( x,y ) - S^{++}_{\alpha \beta} (
x,y )
                                                     \label{eq:sr}
  \end{eqnarray}
and
  \begin{eqnarray}
     S^{\rm A}_{\alpha \beta} ( x,y )
              &=& - \theta(y_{0}-x_{0}) \left\{ S^{+-}_{\alpha \beta} ( x,y
) -
                                 S^{-+}_{\alpha \beta} ( x,y ) \right\}
                                                        \nonumber \\
              &=& S^{--}_{\alpha \beta} ( x,y ) - S^{+-}_{\alpha \beta} (
x,y )
               =  S^{-+}_{\alpha \beta} ( x,y ) - S^{++}_{\alpha \beta} (
x,y ).
                                                         \label{eq:sa}
  \end{eqnarray}
In addition, the following relations are evident:
  \begin{eqnarray}
    S^{\rm R}_{\alpha \beta} ( x,y ) - S^{\rm A}_{\alpha \beta} ( x,y )
       &=& S^{+-}_{\alpha \beta} ( x,y ) - S^{-+}_{\alpha \beta} ( x,y ) \\
    S^{\rm R}_{\alpha \beta} ( x,y ) + S^{\rm A}_{\alpha \beta} ( x,y )
       &=& S^{--}_{\alpha \beta} ( x,y ) - S^{++}_{\alpha \beta} ( x,y ).
  \end{eqnarray}
It is often useful to alternatively consider the matrix of Green functions
  \begin{equation}
    \underline{S}' = \left( \begin{array}{cc}
                                 0       & S^{\rm A} \\
                               S^{\rm R} & S^{\rm K}
                            \end{array} \right)  ,
  \end{equation}
that can be obtained from $\underline S$ by a transformation
 \begin{equation}
    \underline{S'} = \underline{R}^{-1} \underline{S}~~ \underline{R}~,
  \end{equation}
with
  \begin{equation}
    \underline{R}      = \frac{1}{\sqrt{2}} \left( \begin{array}{cc}
                                                      1 & 1 \\
                                                     -1 & 1
                                                   \end{array} \right)~,
    \underline{R}^{-1} = \frac{1}{\sqrt{2}} \left( \begin{array}{cc}
                                                      1 & -1 \\
                                                      1 &  1
                                                   \end{array} \right)~.
  \end{equation}
The transformed self-energy matrix $\underline \Sigma'$ is
\begin{equation}
    \underline{\Sigma}' = \left( \begin{array}{cc}
                               \Sigma^{\rm K} & \Sigma^{\rm R}  \\
                               \Sigma^{\rm A} &      0
                            \end{array} \right)  . 
 \end{equation}
with
\begin{eqnarray}
\Sigma^K &=& \Sigma^{--} + \Sigma^{++} \label{tsk} \\
\Sigma^R &=& \Sigma^{--} + \Sigma^{-+}  \label{tsr}\\
\Sigma^A &=& \Sigma^{--} + \Sigma^{+-}. \label{tsa}
\end{eqnarray}
This can be seen directly, on using the equation
\begin{equation}
\Sigma^{++} + \Sigma^{--} = - (\Sigma^{+-} + \Sigma^{-+}),
\label{relationsig}
\end{equation}
which, in itself can be derived via Eq.(\ref{a3}).    The Dyson equation 
 for the transformed matrix is
\begin{equation}
   \underline{S'}(x,y) = \underline{S'}^{(0)}(x,y)
                  + \int d^{4}z d^{4}w \underline{S'}^{(0)}(x,w)
                                        \underline{\Sigma'}(w,z)
                                        \underline{S'}(z,y)
\label{a16}
  \end{equation}
leading to equations for the retarded and advanced Green functions that are
individually closed:
  \begin{equation}
    S^{\rm R,A}_{\beta \gamma}(x,y) = S^{\rm R,A(0)}_{\beta \gamma}(x,y)
                  + \int d^{4}z d^{4}w S^{\rm R,A(0)}_{\beta\mu}(x,w)
                                       \Sigma^{\rm R,A}_{\mu \nu}(w,z)
                                       S^{\rm R,A}_{\nu \gamma}(z,y)
  \end{equation}
  \begin{equation}
    \left( i \hbar \not\!\partial_{x} -m_0 \right)_{\alpha\beta} S^{\rm R,A}_
{\beta\gamma}(x,y) =
         \delta_{\alpha \gamma} \delta^{4} (x-y)
                  + \int d^{4}z  \Sigma^{\rm R,A}_{\alpha \beta}(x,z)
                                 S^{\rm R,A}_{\beta \gamma}(z,y).
\label{a19}
  \end{equation}
The equation of motion for the Keldysh Green function $S^K$ is found on
multiplying out Eq.(\ref{a16}) and using the fact that $(i\not\!\partial_x
-m_0) S^K{}^{(0)} =0$.     One then has
\begin{equation}
 \left( i \hbar \not\!\partial_{x} -m_0 \right) S^K =
\int(\Sigma^K S^A + \Sigma^R S^K) d^4z, 
\end{equation}
i.e. an integro-differential equation, whose interpretation should also lead
to the 
generalization of the Boltzmann equation \cite{landau}.   This, together with
Eq.(\ref{a19}) form a complete set of equations.   Usually, the solution of
either $S^R$ or $S^A$ is required as these are hermitian conjugates of one
another.

We now write down the solutions for the free particle case.   It is simplest
to
evaluated $S^{R,A}$.    A direct evaluation leads to the form
\begin{equation}
S^{R,A}(p) = \frac{\not p + m_0}{p^2 - m_0^2 \pm i\epsilon p_0}
.
\label{sretfree}
\end{equation}
When a self-energy is included, one has
\begin{equation}
S^{R,A}(p) = \frac{\not p + (m_0 + \Sigma^{R,A})}{p^2 - (m_0 +
\Sigma^{R,A})^2
+i\epsilon p_0}.
\label{sret}
\end{equation}
Using Eqs.(\ref{eq:sr}),
 (\ref{decomp}) and (\ref{decomp2}), one may verify that $S^{--}(X,p)$
and $S^{++}(X,p)$ take the forms that have been given in Eqs.(\ref{smm}) and 
(\ref{spp}).

\section{Wigner transforms}

The transformation rules for applying the Wigner transform to the functions
required in the equations of motion are
\begin{eqnarray}
\partial_y^\mu f(x,y) &\rightarrow& (i\frac{p^\mu}\hbar
 + \frac 12 \partial^\mu_X)f(X,p)
 \\
\partial_x^\mu f(x,y) &\rightarrow& (-i\frac{p^\mu}\hbar
 + \frac 12 \partial^\mu_X)f(X,p)
 \\
f(y)g(x,y)&\rightarrow& f(X)\exp\left(\frac
{i\hbar}2\frac{\overleftarrow\partial}{\partial X^\mu}\frac{\overrightarrow 
\partial}{\partial p_\mu}\right)g(X,p)
 \\
f(x)g(x,y) &\rightarrow& f(X)\exp\left(-\frac
{i\hbar}2\frac{\overleftarrow\partial}{\partial X^\mu}\frac{\overrightarrow    
\partial}{\partial p_\mu}\right) g(X,p) 
 \\
\int d^4z f(x,z) g(z,y) &\rightarrow& f(X,p)\exp \left(-\frac{i\hbar}2
(\frac{\overleftarrow\partial}{\partial X^\mu}\frac {\overrightarrow\partial}
{\partial p_\mu} - \frac{\overleftarrow\partial}{\partial p_\mu}
\frac{\overrightarrow \partial}{\partial X^\mu})\right) g(X,p).
\end{eqnarray}
These relations 
can be derived simply from the definition for the Wigner transform,
Eq.(\ref{wig}) and retaining only first order derivatives in the expansion need
be kept.

\section{Vlasov and constraint equations}

In this Appendix, we derive the Vlasov equation from the Hartree 
approximation to the self-energy in the transport equation, and the
constrained equation in this approximation.    Starting with the transport
equation, Eq.(\ref{hartreetransport}), 
\begin{equation}
\frac {i\hbar}2\{\gamma^\mu,\frac{\partial S^{-+}(X,p)}{\partial X^\mu}\}
+[\not p,S^{-+}(X,p)] =  - i\hbar \partial_\mu \Sigma^A(X)
\partial_p^\mu S^{-+}(X,p),
\label{htagain}
\end{equation}
we insert Eq.(\ref{smp}) for $S^{-+}$ for the Green function.   The 
anticommutator on the left hand side of Eq.(\ref{htagain}) does not 
vanish, whereas the commutator does after a spinor trace is performed. 
  One finds
\begin{eqnarray}
\frac{\partial}{\partial X^\mu}&(&\frac{p^\mu}{2E_p} [\delta(p_0-E_p)
f_q(X,p) - \delta(p_0+E_p)\bar f_{\bar q}(X,-p)]) \nonumber \\
&+& m(X)\partial_\mu m(X)\partial_p^\mu(\frac 1{2E_p}[\delta
(p_0-E_p)f_q(X,p) - \delta(p_0+E_p)\bar f_{\bar q}(X,-p)]) = 0 
\label{aa2}
\end{eqnarray}
where a spinor trace has been performed, and we have made use of the fact
that $\partial_\mu\Sigma^A(X) = \partial_\mu m(X)$, since $m_0$ is a 
constant.   Now this equation is integrated over a positive energy range.
In this fashion, only the quark distribution function can survive.  One
thus has
\begin{eqnarray}
\frac{\partial}{\partial X^\mu}\int_{\Delta_+} dp_0
\frac{p^\mu}{E_p}& &\delta(p_0-E_p) f_q(X,p) \nonumber \\
& +& m(X)\partial_\mu m(X)
\int_{\Delta_+} dp_0\partial_p^\mu\frac 1{E_p}\delta(p_0-E_p) f_q(X,p)
=0.
\label{aa3}
\end{eqnarray}
It is useful to write the Lorentz indices out explicitly.   Using the
notation $i=1..3$, this reads
\begin{equation}
\frac{\partial}{\partial X^0} f_q(X,\vec p) + p_i\partial^i(\frac{
f_q(X,\vec p)}{E_p}) + m(X)\partial_i m(X)\partial_p^i(\frac{f_q(X,\vec p)}
{E_p}) = 0.
\label{a4}
\end{equation}
Note that no term proportional to $m(X)\partial_0 m(X)$ occurs, as this
term contains a vanishing surface integral.   If one explicitly performs
the differentiations indicated, noting that $\vec\nabla_xE_p^{-1} =
-(m(X)\vec\nabla m(X))/E_p^3$ and $\vec\nabla_pE_p^{-1} = -\vec p/E_p^3$,
one finds
\begin{equation}
\frac{\partial}{\partial X^0} f_q(X,\vec p) + \vec v\cdot \vec\nabla f_q
(X,\vec p) - \vec\nabla E_p\cdot\vec\nabla_p f_q(X,\vec p) = 0,
\label{a5}
\end{equation}
where we have written $\vec v = \vec p/E_p$.    This is the Vlasov equation,
and it is manifestly non-covariant.   One can, however, extract a factor
$E_p^{-1}$ from the above equation, and write
\begin{equation}
\frac 1{E_p}(p^\mu\partial_\mu f_q(X,\vec p) + m(X)\partial_\mu m(X)\partial
_p^\mu f_q(X,\vec p)) = 0,
\label{a6}
\end{equation}
giving it a covariant appearance.  Of course, in the above expression,
$p^0=E_p$, while there is no $p_0$ dependence in the distribution function.
It is this equation that has appeared as Eq.(\ref{vlasov}) in the main text. 

The derivation of the constraint equation is far simpler.    Consider
Eq.(\ref{hartreeconstraint}), and insert the quasiparticle form
Eq.(\ref{smp}) for $S^{-+}$.     The commutator, which is the first
term of this equation, gives no contribution upon taking the spinor
trace.   The remaining anticomutator terms can be combined, to give
\begin{equation}
{\rm tr} \{\not p - m(X), S^{-+} \} = 0
\label{a7}
\end{equation}
since $m(X) = m_0 + \Sigma^A(X) = m_0 + \Sigma^R(X)$ in this approximation.
Integrating on $\Delta_+$ leads directly to the equation
\begin{equation}
(p^2-m^2(X))f_q(X,p) = 0
\label{a8}
\end{equation}
that was quoted in Eq.(\ref{hatconstr}).

\section{Pseudoscalar collision term}

In this appendix, we provide the detail required to construct the collision
term from Eq.(\ref{picoll}) with the self-energy as calculated via Fig.4
using the pseudoscalar vertices $i\gamma_5$.    The self-energy has
been given in Eq.(\ref{sigpi}).    On inserting this into
Eq.(\ref{picoll}) 
 and
multiplying out the result, one again finds {\it inter alia} off-shell
processes of the form shown in Fig.~6, where the interaction vertex is now
pseudoscalar.    Retaining only those terms that survive energy-momentum
conservation, and performing the same manipulations that led to
Eq.(\ref{monster}), one finds the analogous result
\begin{eqnarray}
J_{{\rm coll},\pi}^{q,{\rm loss}} &=& -\frac{\pi i}{E_p} \frac{4G^2
i^3}\hbar
\int\frac{d^3p_1}{(2\pi\hbar)^3 2E_{p_1}} \frac{d^3p_2}{(2\pi\hbar)^3 2E_{p_2}}
\frac{d^3p_3}{(2\pi\hbar)^3 2E_{p_3}} (2\pi\hbar)^4 \nonumber \\
&\times& \{\delta^4(p-p_1+p_2-p_3) [{\rm tr}(u\bar u)_p\gamma_5
(u\bar u)_1\gamma_5{\rm tr}
(u\bar u)_2 \gamma_5
(u\bar u)_3 \gamma_5\nonumber \\
& & \quad - {\rm tr} (u\bar u)_p\gamma_5(u\bar u)_1\gamma_5(u\bar u)_2
\gamma_5(u\bar u_3)\gamma_5]
\bar f_q(p_1) f_q(p_2)\bar f_q(p_3) f_q(p)
\nonumber \\
&+& \delta^4(p-p_1-p_2 +p_3)[{\rm tr}(u\bar u)_p(u\bar u)_1 {\rm tr} (v\bar
v)_2
(v\bar v)_3 \nonumber \\
& & \quad - {\rm tr}(u\bar u)_p\gamma_5(u\bar u)_1\gamma_5 (v\bar
v)_2\gamma_5(v\bar v)_3\gamma_5]
\bar f_q(p_1)\bar f_{\bar q}(p_2)f_{\bar q}(p_3)f_q(p)
\nonumber \\
&+& \delta^4(p+p_1-p_2-p_3)[{\rm tr}(u\bar u)_p\gamma_5(v\bar v)_1\gamma_5
{\rm tr} (v\bar
v)_2\gamma_5
(u\bar u)_3\gamma_5 \nonumber \\
& & \quad  - {\rm tr}(u\bar u)_p\gamma_5(v\bar v)_1\gamma_5(v\bar v)_2
\gamma_5(u\bar u)_3\gamma_5]
f_{\bar q}(p_1)\bar f_{\bar q}(p_2)\bar f_{q}(p_3) f_q(p)\}.
\nonumber
\\
\label{monsterpi}
\end{eqnarray}
Note that the terms with two traces arose from the direct graph Fig.~4(a)
while the term with one trace originated from the exchange graph Fig.~4(b).
now, once again, Fig.~7 may be examined with pseudoscalar vertices, leading to
the $t$ and $s$ channel forms,
\begin{eqnarray}
-iM^t_\pi(qq\rightarrow qq) &=& \frac{2iG}\hbar
\bar u_1^\alpha(i\gamma_5)^{\alpha\beta}
u_p^\beta
\bar u_3^{\alpha'}(i\gamma^5)^{\alpha'\beta'}u_2^{\beta'} \label{mqq2} \\
-iM^u_\pi(qq\rightarrow qq) &=& \frac{2iG}\hbar\bar u_1^\alpha(i\gamma_5)^{\alpha\beta}
u_2^\beta\bar u_3^{\alpha'} (i\gamma_5)^{\alpha'\beta'} u_p^{\beta'}
\label{mqq3}
\end{eqnarray}
for quark-quark scattering.   The squares of these terms correspond to the
first two terms of Eq.(\ref{monsterpi}), while the latter two terms are given
by the combination
$\sum (M^u_\pi M^t_\pi{}^* + M^t_\pi M^u_\pi{}^*)$.   In a similar
fashion,
the quark-antiquark channels from Fig.~8 are
\begin{eqnarray}
-iM_\pi^s(q\bar q\rightarrow q\bar q) &=& \frac{2iG}\hbar
\bar u_1^\alpha(i\gamma_5)^
{\alpha\beta} v_2^\beta\bar
v_3^{\alpha'} (i\gamma_5)^{\alpha'\beta'}u_p^{\beta'}
\\
-iM_\pi^t(q\bar q\rightarrow q\bar q) &=& \frac{2iG}\hbar \bar u_1^\alpha
(i\gamma_5)^{\alpha\beta}
 u_p^\beta \bar
v_3^{\alpha'}(i\gamma_5)^{\alpha'\beta'} v_2^{\beta'}
\label{mqqbpi}
\end{eqnarray}
which combine in the same way to yield the terms that form the sum
multiplying the second $\delta$-function.   One thus has
\begin{eqnarray}
J_q^{{\rm loss},\pi} = &-& \frac{\pi\hbar}{E_p} \int\frac{d^3p_1}{(2\pi\hbar)^3
2E_{p_1}} \frac{d^3p_2}{(2\pi\hbar)^3 2E_{p_2}}
\frac{d^3p_3}{(2\pi\hbar)^3 2E_{p_3}} (2\pi\hbar)^4  \nonumber \\
&\times&\{\delta^4(p-p_1+p_2-p_3)\frac 12\sum |M^\pi_{qq\rightarrow qq}
(p2\rightarrow 13)|^2\bar f_q(p_1)f_q(p_2)\bar f_q(p_3)f_q(p) \nonumber \\
&+&\delta^4(p-p_1-p_2+p_3)\sum |M^\pi_{q\bar q\rightarrow q\bar q}
(p3
\rightarrow 12)|^2\bar f_q(p_1)\bar f_{\bar q}(p_2)f_{\bar q}(p_3)f_q(p)\},
\nonumber \\
\label{matrpi}
\end{eqnarray}
where $|M^\pi_{qq\rightarrow qq}| = |M_\pi^t(qq\rightarrow qq) -
M_\pi^u(qq\rightarrow qq)|$ and $|M^\pi_{q\bar q\rightarrow q\bar q}|
= |M_\pi^s(q\bar q\rightarrow q\bar q) + M_\pi^t(q\bar q\rightarrow q\bar
q)|$
again contain the relevant $s$, $t$ and $u$ channels.
This was quoted as Eq.(\ref{matrpi2}).

\section{Mixed collision term}

In this appendix, the contribution of $\Sigma_{{\rm mixed}}^{+-}$ to the
collision term is examined.
As before, $S^{-+}$ and $S^{+-}$ are inserted into $\Sigma_{{\rm
mixed}}^{+-}$ given by Eq.(\ref{sigmixed}), and the expression 
for $J^{q,{\rm loss}}_{{\rm coll, mixed}}$ of
Eq.(\ref{lossmix}) is multiplied out.
One may directly identify the energy-momentum conserving terms
in this case to be
\begin{eqnarray}
J^{q,{\rm loss}}_{{\rm coll, mixed}} &=& -\frac{\pi i}{E_p} \frac{4G^2 i^3}
\hbar
\int\frac{d^3p_1}{(2\pi\hbar)^3 2E_{p_1}} \frac {d^3p_2}
{(2\pi\hbar)^3 2E_{p_2}}\frac{d^3p_3}{(2\pi\hbar)^3 2E_{p_3}} (2\pi\hbar)^4  \nonumber \\
&\times&\{\delta(p-p_1+p_2-p_3)[{\rm tr} (u\bar u)_p\gamma_5(u\bar
u)_1(u\bar u)_2\gamma_5(u\bar u)_3 \nonumber \\
& & \quad + {\rm tr}(u\bar u)_p(u\bar u)_1\gamma_5(u\bar u)_2(u\bar u)_3
\gamma_5]\bar f_q(p_1) f_2(p_2)\bar f_q(p_3) f_q(p) \nonumber \\
&+&\delta(p-p_1-p_2+p_3)[{\rm tr}(u\bar u)_p\gamma_5(u\bar u)_1(v\bar
v)_2\gamma_5(v\bar v)_3 \nonumber \\
& & + {\rm tr}(u\bar u)_p(u\bar u)_1\gamma_5(v\bar v)_2(v\bar v)_3\gamma_5
+ {\rm tr} (u\bar u)_p\gamma_5(v\bar v)_3(v\bar v)_2\gamma_5(u\bar u)_1
\nonumber \\
& & +{\rm tr} (u\bar u)_p(v\bar v)_3\gamma_5(v\bar v)_2(u\bar u)_1\gamma_5]
\bar f_q(p_1)\bar f_{\bar q}(p_2) f_{\bar q}(p_3) f_q(p) \}.
\label{awful}   
\end{eqnarray}
A slight simplification can be introduced on letting $2\leftrightarrow 3$ in
the second term.   One has
\begin{eqnarray}
J^{q,{\rm loss}}_{{\rm coll, mixed}} &=& -\frac{\pi i}{E_p} \frac{4G^2
i^3}\hbar
\int\frac{d^3p_1}{(2\pi\hbar)^3 2E_{p_1}} \frac {d^3p_2}
{(2\pi\hbar)^3 2E_{p_2}}\frac{d^3p_3}{(2\pi\hbar)^3 2E_{p_3}}
\nonumber \\
&\times& (2\pi\hbar)^4
\delta(p-p_1+p_2-p_3)
\nonumber \\
&\times &[{\rm tr} (u\bar u)_p\gamma_5(u\bar
u)_1(u\bar u)_2\gamma_5(u\bar u)_3
 + {\rm tr}(u\bar u)_p(u\bar u)_1\gamma_5(u\bar u)_2(u\bar u)_3
\gamma_5] \nonumber \\
& & \quad \times \bar f_q(p_1) f_2(p_2)\bar f_q(p_3) f_q(p) \nonumber \\
&+&[{\rm tr}(u\bar u)_p\gamma_5(u\bar u)_1(v\bar
v)_2\gamma_5(v\bar v)_3
 + {\rm tr}(u\bar u)_p(u\bar u)_1\gamma_5(v\bar v)_2(v\bar v)_3\gamma_5
\nonumber \\
& & + {\rm tr} (u\bar u)_p\gamma_5(v\bar v)_3(v\bar v)_2\gamma_5(u\bar u)_1
 +{\rm tr} (u\bar u)_p(v\bar v)_3\gamma_5(v\bar v)_2(u\bar u)_1\gamma_5]
\nonumber \\
& & \quad \times \bar f_q(p_1)\bar f_{\bar q}(p_2) f_{\bar q}(p_3) f_q(p)
\}.
\label{awful2}
\end{eqnarray}
It is now essential to identify these terms.  In order to do so, we note
that,   for the scattering processes
of Fig.~7 that include both $t$ and $u$ channels, the complete scattering
amplitude is
\begin{equation}
M_{qq\rightarrow qq} = M_\sigma^t - M_\sigma^u + M^t_\pi - M_\pi^u.
\label{ems}
\end{equation}
Squaring this leads to
\begin{eqnarray}
|M_{qq\rightarrow qq}|^2 &=& |M^\sigma|^2 + |M^\pi|^2 +
M_\sigma^tM_\pi^t{}^*
- M_\sigma^tM_\pi^u{}^* \nonumber \\
&-& M_\sigma^uM_\pi^t{}^* + M_\sigma^uM_\pi^u{}^* + M_\pi^tM_\sigma^t{}^*
- M_\pi^tM_\sigma^u{}^* \nonumber \\
&-& M_\pi^uM_\sigma^t{}^* + M_\pi^uM_\sigma^u{}^*.
\label{sq}
\end{eqnarray}
The first two terms of this expression have already been generated in
Eq.(\ref{sigmapi}).   Performing the sum over spinor indices, one
sees from Eqs.(\ref{mqq1}), (\ref{mqq}), (\ref{mqq2}) and (\ref{mqq3}),
that
\begin{equation}
\sum M_\sigma^t M_\pi^t{}^* = 0 = \sum M_\sigma^u M_\pi^u{}^*,
\label{simple}
\end{equation}
but that 
\begin{eqnarray}
\sum M_\sigma^u M_\pi^t{}^* &=& -\frac{4G^2}{\hbar^2}
{\rm tr} (u\bar u)_p\gamma_5
(u\bar u)_1(u\bar u)_2\gamma_5 (u\bar u)_3 \nonumber \\
\sum M_\sigma^t M_\pi^u{}^* &=& - \frac{4G^2}{\hbar^2}
{\rm tr} (u\bar u)_p\gamma_5(u\bar
u)_3
(u\bar u)_2\gamma_5 (u\bar u)_1 \nonumber \\
\sum M_\pi^tM_\sigma^u{}^* &=& - \frac{4G^2}{\hbar^2}
{\rm tr} (u\bar u)_p (u\bar
u)_3\gamma_5(u\bar u)_2 (u\bar u)_1\gamma_5 \nonumber\\
\sum M_\pi^u M_\sigma^t{}^* &=& -\frac{4G^2}{\hbar}{\rm tr} (u\bar u)_p (u\bar u)_1
\gamma_5(u\bar u)_2(u\bar u)_3\gamma_5
\label{terms}
\end{eqnarray}
Now, direct inspection of Eq.(\ref{awful})
 indicates that all the terms of Eq.(\ref{terms})
are present, if one extracts a factor $1/2$ and makes use of the symmetry
in the variables 1 and 3.   The $q\bar q$ scattering term can be viewed
similarly in its $s$ and $t$ channels for $\pi$ and $\sigma$, and the 
terms may all be identified.   One thus finds the result for $J^{q,{\rm loss}}
_{{\rm coll, mixed}}$ that was quoted in Eq.(\ref{wonder}).

\newpage
\begin{center}
FIGURE CAPTIONS.
\end{center} 

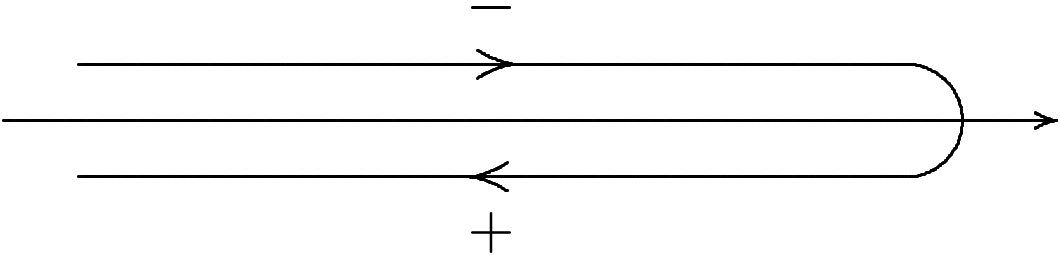
Fig.1 :   Closed time path on which the Green functions are defined.

\vskip 0.1in

Fig.2 :   Diagrammatic expansion of the Dyson equation for the Green
function $S^{++}$.

\vskip 0.1in
Fig.3 :   Self-consistent Hartree approximation to the self-energy.  

\vskip 0.1in
Fig.4 :   Direct (a) and exchange (b) graphs that contribute to 
$\Sigma^{-+}$ and which contain two interaction lines.

\vskip 0.1in
Fig.5 :  Mixed graphs that contribute to $\Sigma^{+-}_{{\rm mixed}}$ and
which contain two interaction lines.

\vskip 0.1in
Fig.6 :  Processes of the form $q\rightarrow (q\bar q)q$ and $(q\bar q)q
\rightarrow q$  as well as vacuum fluctuations
that are suppressed in the quasiparticle approximation.
These are heuristic graphs and are not Feynman graphs.   They should be
read from left to right.

\vskip 0.1in
Fig.7 :  The $t$ and $u$ channel Feynman 
graphs for elastic quark-quark scattering.

\vskip 0.1in
Fig.8 :  The $s$ and $t$ channel Feynman graphs for elastic  quark-antiquark
scattering.

\vfill

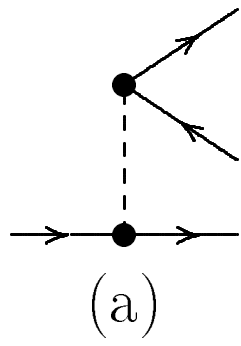
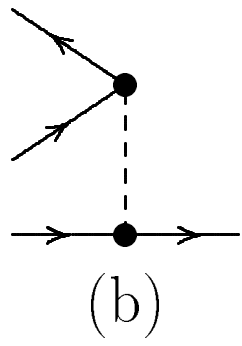
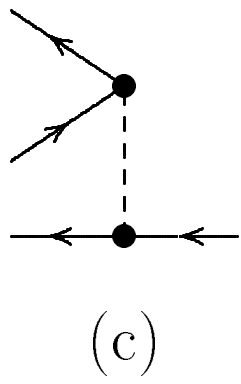
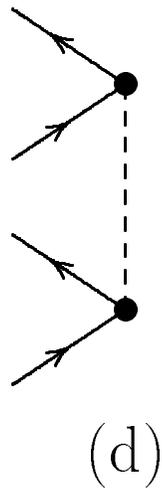
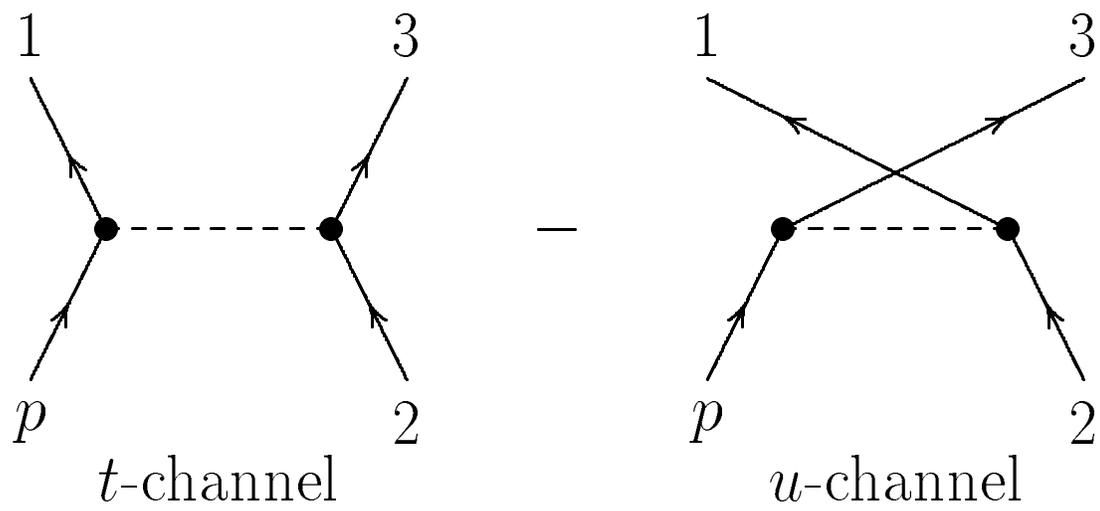
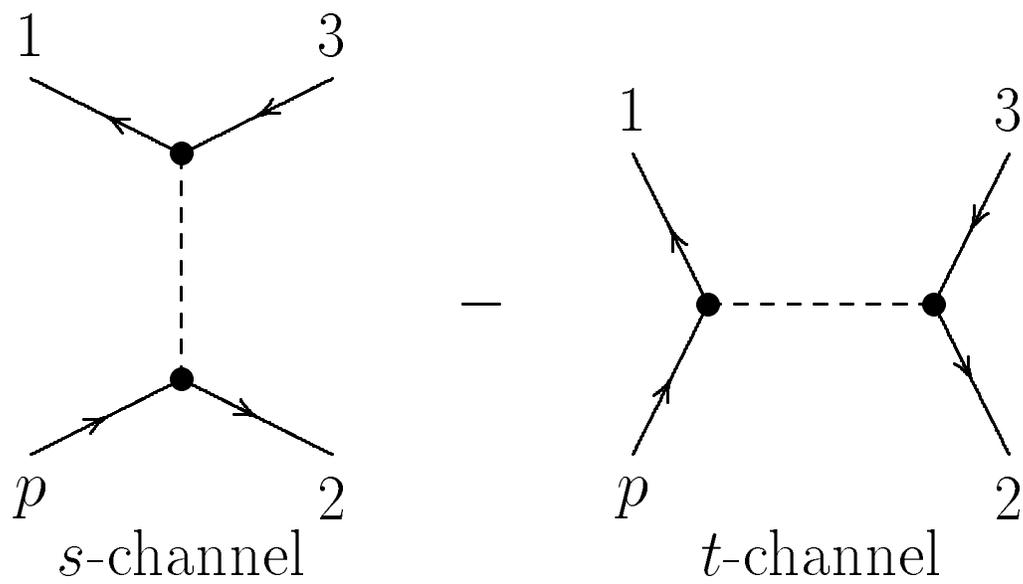
\end{document}